\newcount\mgnf\newcount\tipi\newcount\tipoformule
\newcount\aux\newcount\aaux\newcount\pie
\mgnf=0          
\tipoformule=0   
\aux=0           
           \def\9#1{\ifnum\aux=1#1\else\relax\fi}
\aaux=0          
\ifnum\mgnf=0
   \magnification=\magstep0
\hsize=13.5truecm\vsize=21truecm
   \parindent=4.pt\fi
\ifnum\mgnf=1
   \magnification=\magstep1
\hsize=16.0truecm\vsize=22.5truecm\baselineskip14pt\vglue3.1truecm
   \parindent=4.pt\fi
%
\let\a=\alpha \let\b=\beta      \let\d=\delta \let\e=\varepsilon
\let\z=\zeta  \let\h=\eta   \let\th=\vartheta\let\k=\kappa \let\l=\lambda
\let\m=\mu    \let\n=\nu           \let\p=\pi    
\let\s=\sigma \let\t=\tau    \let\ch=\chi
\let\ps=\psi  \let\o=\omega 
     \let\L=\Lambda
            \let\O=\Omega

{\count255=\time\divide\count255 by 60 \xdef\oramin{\number\count255}
        \multiply\count255 by-60\advance\count255 by\time
   \xdef\oramin{\oramin:\ifnum\count255<10 0\fi\the\count255}}
\def\ora{\oramin }
 
\def\data{\number\day/\ifcase\month\or gennaio \or febbraio \or marzo \or
aprile \or maggio \or giugno \or luglio \or agosto \or settembre
\or ottobre \or novembre \or dicembre \fi/\number\year;\ \ora}
 
\setbox200\hbox{$\scriptscriptstyle \data $}
 
\newcount\pgn \pgn=1
\def\foglio{\number\numsec:\number\pgn
\global\advance\pgn by 1}
\def\foglioa{A\number\numsec:\number\pgn
\global\advance\pgn by 1}
 
 
\global\newcount\numsec\global\newcount\numfor
\global\newcount\numfig
\gdef\profonditastruttura{\dp\strutbox}
\def\senondefinito#1{\expandafter\ifx\csname#1\endcsname\relax}
\def\SIA #1,#2,#3 {\senondefinito{#1#2}
\expandafter\xdef\csname #1#2\endcsname{#3} \else
\write16{???? ma #1,#2 e' gia' stato definito !!!!} \fi}
\def\etichetta(#1){(\veroparagrafo.\veraformula)
\SIA e,#1,(\veroparagrafo.\veraformula)
 \global\advance\numfor by 1
\9{\write15{\string\FU (#1){\equ(#1)}}}
\9{ \write16{ EQ \equ(#1) == #1  }}}
\def \FU(#1)#2{\SIA fu,#1,#2 }
\def\etichettaa(#1){(A\veroparagrafo.\veraformula)
 \SIA e,#1,(A\veroparagrafo.\veraformula)
 \global\advance\numfor by 1
\9{\write15{\string\FU (#1){\equ(#1)}}}
\9{ \write16{ EQ \equ(#1) == #1  }}}
\def\getichetta(#1){Fig. \verafigura
 \SIA e,#1,{\verafigura}
 \global\advance\numfig by 1
\9{\write15{\string\FU (#1){\equ(#1)}}}
\9{ \write16{ Fig. \equ(#1) ha simbolo  #1  }}}
\newdimen\gwidth
\def\BOZZA{
\def\alato(##1){
 {\vtop to \profonditastruttura{\baselineskip
 \profonditastruttura\vss
 \rlap{\kern-\hsize\kern-1.2truecm{$\scriptstyle##1$}}}}}
\def\galato(##1){ \gwidth=\hsize \divide\gwidth by 2
 {\vtop to \profonditastruttura{\baselineskip
 \profonditastruttura\vss
 \rlap{\kern-\gwidth\kern-1.2truecm{$\scriptstyle##1$}}}}}
\footline={\rlap{\hbox{\copy200}\ $\st[\number\pageno]$}\hss\tenrm
\foglio\hss}}
\def\alato(#1){}
\def\galato(#1){}
\def\veroparagrafo{\number\numsec}\def\veraformula{\number\numfor}
\def\verafigura{\number\numfig}
\def\geq(#1){\getichetta(#1)\galato(#1)}
\def\Eq(#1){\eqno{\etichetta(#1)\alato(#1)}}
\def\eq(#1){\etichetta(#1)\alato(#1)}
\def\Eqa(#1){\eqno{\etichettaa(#1)\alato(#1)}}
\def\eqa(#1){\etichettaa(#1)\alato(#1)}
\def\eqv(#1){\senondefinito{fu#1}$\clubsuit$#1\write16{Manca #1 !}%
\else\csname fu#1\endcsname\fi}
\def\equ(#1){\senondefinito{e#1}\eqv(#1)\else\csname e#1\endcsname\fi}
 
\openin13=#1.aux \ifeof13 \relax \else
\input #1.aux \closein13\fi
\openin14=\jobname.aux \ifeof14 \relax \else
\input \jobname.aux \closein14 \fi
\9{\openout15=\jobname.aux}
\newskip\ttglue

\font\ottorm=cmr8\font\ottoi=cmmi8\font\ottosy=cmsy8
\font\ottobf=cmbx8\font\ottott=cmtt8
\font\ottosl=cmsl8\font\ottoit=cmti8
\font\sixrm=cmr6\font\sixbf=cmbx6\font\sixi=cmmi6\font\sixsy=cmsy6
\font\fiverm=cmr5\font\fivesy=cmsy5\font\fivei=cmmi5\font\fivebf=cmbx5
\def\ottopunti{\def\rm{\fam0\ottorm}
\textfont0=\ottorm \scriptfont0=\sixrm \scriptscriptfont0=\fiverm
\textfont1=\ottoi \scriptfont1=\sixi   \scriptscriptfont1=\fivei
\textfont2=\ottosy \scriptfont2=\sixsy   \scriptscriptfont2=\fivesy
\textfont3=\tenex \scriptfont3=\tenex   \scriptscriptfont3=\tenex
\textfont\itfam=\ottoit  \def\it{\fam\itfam\ottoit}%
\textfont\slfam=\ottosl  \def\sl{\fam\slfam\ottosl}%
\textfont\ttfam=\ottott  \def\tt{\fam\ttfam\ottott}%
\textfont\bffam=\ottobf  \scriptfont\bffam=\sixbf
\scriptscriptfont\bffam=\fivebf  \def\bf{\fam\bffam\ottobf}%
\tt \ttglue=.5em plus.25em minus.15em
\setbox\strutbox=\hbox{\vrule height7pt depth2pt width0pt}%
\normalbaselineskip=9pt\let\sc=\sixrm \normalbaselines\rm}
\catcode`@=11      
\def\footnote#1{\edef\@sf{\spacefactor\the\spacefactor}#1\@sf
\insert\footins\bgroup\ottopunti
 \interlinepenalty100 \let\par=\endgraf
   \leftskip=0pt \rightskip=0pt
   \splittopskip=10pt plus 1pt minus 1pt \floatingpenalty=20000
   \smallskip\item{#1}\bgroup\strut\aftergroup\@foot\let\next}
\skip\footins=12pt plus 2pt minus 4pt
\dimen\footins=30pc
\catcode`@=12
%

%
%
%
\newdimen\xshift \newdimen\xwidth \newdimen\yshift
 
\def\ins#1#2#3{\vbox to0pt{\kern-#2\hbox{\kern#1 #3}\vss}\nointerlineskip}
 
\def\eqfig#1#2#3#4#5{
\par\xwidth=#1 \xshift=\hsize \advance\xshift
by-\xwidth \divide\xshift by 2
\yshift=#2 \divide\yshift by 2
\line{\hglue\xshift \vbox to #2{\vfil
#3 \includegraphics{#4.ps}
}\hfill\raise\yshift\hbox{#5}}}
 
 
\def\8{\write13}  


\def\V#1{{\bf#1}}
\def\T#1{#1\kern-4pt\lower9pt\hbox{$\widetilde{}$}\kern4pt{}}
\let\dpr=\partial
\let\io=\infty\let\ig=\int
\def\fra#1#2{{#1\over#2}}\let\0=\noindent
 
\def\guida{\leaders\hbox to 1em{\hss.\hss}\hfill}
\def\tende#1{\vtop{\ialign{##\crcr\rightarrowfill\crcr
              \noalign{\kern-1pt\nointerlineskip}
              \hglue3.pt${\scriptstyle #1}$\hglue3.pt\crcr}}}
\def\otto{{\kern-1.truept\leftarrow\kern-5.truept\to\kern-1.truept}}

\def\pagina{\vfill\eject}

\def\tst{\textstyle}\def\st{\scriptscriptstyle}
 
\ifnum\mgnf=0
\def\*{\vskip0.3truecm}
\fi
\ifnum\mgnf=1
\def\*{\vskip0.5truecm}
\fi


\def\ap{\hbox{\it a priori\ }}
\def\ie{\hbox{\it i.e.\ }}
\def\fiat{{}}
\let\ciao=\bye
 
\def\\{\hfill\break}
\def\={{ \; \equiv \; }}
\def\annota#1#2{\footnote{${}^#1$}{{\parindent=0pt
\0#2\vfill}}}
 
\ifnum\aux=1\BOZZA\else\relax\fi
\ifnum\tipoformule=0\let\Eq=\eqno\def\eq{}\let\Eqa=\eqno\def\eqa{}
\def\equ{{}}\footline={\hss\tenrm\folio\hss}\fi
\ifnum\aaux=0\let\foglio=\folio\else\relax\fi
 

\def\TT{{\cal T}}   \def\BB{{\cal B}}   
\def\PP{{\cal P}}   \def\HHH{{\cal H}}  \def\RR{{\cal R}}
  \def\VVV{{\cal V}}  \def\QQ{{\cal Q}}
  \def\OO{{\cal O}}   \def\UU{{\cal U}}
\def\CC{{\cal C}}
 
\def\Val{{\,{\rm Val}}}
 
\font\titolo=cmbx12
\font\titolone=cmbx12 scaled \magstep2
\font\cs=cmcsc10
\font\ss=cmss10
\font\sss=cmss8
\font\crs=cmbx8
 
\font\tenmib=cmmib10
\font\sevenmib=cmmib10 scaled 800
 
\textfont5=\tenmib  \scriptfont5=\sevenmib  \scriptscriptfont5=\fivei
 
\def\HH{{\V H}} \def\hh{{\V h}}
\def\AA{{\V A}} \def\NN{{\V N}}
\def\ff{{\V f}} \def\aaa{{\V a}}
\def\ee{{\V e}} \def\mm{{\V m}}
\def\xx{{\V x}} \def\yy{{\V y}} \def\gg{{\V g}}
\mathchardef\aa = "050B
\mathchardef\bb = "050C
\mathchardef\xxx= "0518
\mathchardef\zz = "0510
\mathchardef\oo = "0521
\mathchardef\ll = "0515
\mathchardef\mm = "0516
\mathchardef\Dp = "0540
\mathchardef\H  = "0548
\mathchardef\FFF= "0546
\mathchardef\ppp= "0570
\mathchardef\nn = "0517
\mathchardef\pps= "0520

\def\nnx{{\{\nn_x\}}}

 
\ifnum\mgnf=0
\def\ZZZ{\hbox{{\ss Z}\kern-3.3pt{\ss Z}}}
\def\zzz{\hbox{{\sss Z}\kern-3.3pt{\sss Z}}}
\fi
\ifnum\mgnf=1
\def\ZZZ{\hbox{{\ss Z}\kern-3.6pt{\ss Z}}}
\def\zzz{\hbox{{\sss Z}\kern-3.6pt{\sss Z}}}
\fi
 
\ifnum\mgnf=0
\def\TTT{\hbox{{\ss T}\kern-5.0pt{\ss T}}}
\def\ttt{\hbox{{\sss T}\kern-4.2pt{\sss T}}}
\fi
\ifnum\mgnf=1
\def\TTT{\hbox{{\ss T}\kern-5.3pt{\ss T}}}
\def\ttt{\hbox{{\sss T}\kern-3.9pt{\sss T}}}
\fi
 
\ifnum\mgnf=0
\def\RRR{\hbox{I\kern-1.4pt{\ss R}}}
\def\rrr{\hbox{{\crs I}\kern-1.3pt{\sss R}}}
\fi
\ifnum\mgnf=1
\def\RRR{\hbox{I\kern-1.7pt{\ss R}}}
\def\rrr{\hbox{{\crs I}\kern-1.6pt{\sss R}}}
\fi
 
\ifnum\mgnf=0
\def\openone{\leavevmode\hbox{\ninerm 1\kern-3.3pt\tenrm1}}%
\fi
\ifnum\mgnf=1
\def\openone{\leavevmode\hbox{\ninerm 1\kern-3.63pt\tenrm1}}%
\fi
\def\1{\ifnum\mgnf=0\pagina\else\relax\fi}
 
\fiat
\vglue0.5truecm
\centerline{\titolone Quasi linear flows on tori: regularity}
\centerline{\titolone of their linearization}
\vskip1.truecm
\centerline{\titolo F. Bonetto$^{\bullet}$, G. Gallavotti$^{*}$,
G. Gentile$^{\dagger}$, V. Mastropietro$^{\star}$}
\vskip.5truecm
\centerline{\ottott $^{\bullet}$ Matematica, Universit\`a di Roma, P.le
Moro 2, 00185 Roma}
 
\centerline{\ottott $^{*}$ Fisica, Universit\`a di Roma, P.le Moro 2,
00185 Roma}
 
\centerline{\ottott $^{\dagger}$ IHES, 35 Route de Chartres, 91440
Bures sur Yvette}
 
\centerline{\ottott $^{\star}$ Matematica, Universit\`a di Roma II,
Via Ricerca Scientifica, 00133 Roma}
 
\vskip1.truecm
 
\line{\vtop{
\line{\hskip1.5truecm\vbox{\advance \hsize by -3.1 truecm
\\{\cs Abstract.}
{\it Under suitable conditions a flow on a torus $C^{(p)}$--close, with
$p$ large enough, to a quasi periodic diophantine rotation is shown to
be conjugated to the quasi periodic rotation by a map that is analytic
in the perturbation size. This result is parallel to Moser's theorem
stating conjugability in class $C^{(p')}$ for some $p'<p$. The extra
conditions restrict the class of perturbations that are allowed.}}
\hfill} }}

\vskip1.2truecm

\centerline{\titolo 1. Introduction}
\*\numsec=1\numfor=1
 
\0{\bf 1.1.} The perturbation of the Hamiltonian of a system
of $\ell$ harmonic oscillators with frequencies
$\fra1{2\p}(\o_{01},\ldots,\o_{0\ell})$
is described by the Hamiltonian
$$\HHH_0(\AA,\aa)=\oo_0\cdot\AA+\e \AA\cdot
\ff(\aa)+ \e \, \AA \cdot F(\AA,\aa)\,\AA \; , \Eq(1.1)$$
where $\AA,\aa\in \RRR^\ell\times \TTT^\ell$ are the {\it action--angle}
variables of the oscillators, $\cdot$ denotes the
scalar product, $\ff$ is a vector and $F$ a matrix that
describe the perturbation structure and $\e$ is the {\it intensity} of
the perturbation.
 
The Hamiltonian system \equ(1.1) is not integrable in general (see for
instance (4.10) in [G3]). Nevertheless, if the {\it unperturbed rotation
vector $\oo_0$} of the oscillators verifies a {\sl diophantine condition}
and if the perturbation is analytic, it is possible to add to the
Hamiltonian a suitable ``{\it counterterm}'' $\V A\cdot \V N_\e(\V A)$,
analytic in the perturbative parameter $\e$ and depending only on the
action variables $\V A$, so that the modified Hamiltonian $\HHH_0+\V A\cdot
\V N_\e(\V A)$ is integrable.
 
This was conjectured in [G1] and proven in [E2], then in [GM2] with the
techniques introduced in [G4,GM1], by exploiting the cancellation
mechanisms operating order by order in the perturbative series defining
the counterterm and the solution of the equations of motion for the
modified Hamiltonian; a third method is in [EV].
 
Note that integrability of the problem \equ(1.1) with $F=0$ is equivalent
to the problem of linearizability of the flow on the torus generated by
the differential equations
$${d\aa\over dt} =\oo_0+\e \ff(\aa) \; , \Eq(1.2)$$
\ie the problem of finding a change of coordinates,
$\aa=\pps+\hh_\e(\pps)$, on the torus $\TTT^\ell$ such that the
equations \equ(1.2) become the trivial quasi periodic linear flow
$d\pps/dt=\oo_0$.
 
A previous partial result about the existence of the counterterm
$N_\e(\V A)$ is in [DS,R,PF], where the one-dimensional Schr\"odinger
equation with a quasi periodic potential is studied: the latter problem
can be shown to be equivalent to the problem of the existence of a
coun\-ter\-term which makes integrable the Hamiltonian of a system of
interacting oscillators, provided $F=0$ and $\ff$ has a very special
form, (see [G2] and \S 5 below): restricted to this case, the proof of
existence of a counterterm making integrable the Hamiltonian follows
also from the analysis of [M1], (see Appendix A4 below).
 
But the ``beginning'' of the interest in the above problems goes back to a
problem similar to \equ(1.2) investigated by Moser in [M1]. There more
general non linear ordinary differential equations are perturbatively
studied under the hypothesis that the {\sl characteristic number}
(defining the linear part of the equations) verifies a {\sl generalized
diophantine condition} (see Appendix A4 for details): under the
assumption that $\ff$ is analytic it is proved that one can add to the
equations counterterms depending (analytically) {\it
only} on the perturbative parameter $\e$ and that the equations so
modified admit a solution analytic in $\e$ for $\e$ small.

The conjecture in [G1] envisages the possibility of introducing a
counterterm depending (analytically) {\it also} on the actions in the
Hamiltonian in such a way that the equations of motion admit an analytic
solution: the two problems (the one studied by Moser and the one studied
in [G2]) deal in general with different equations.  Nevertheless, if the
perturbation is linear in the action variables in
\equ(1.1), \ie $F=0$, then the counterterm $\NN_\e(\AA)$ turns out to be
$\AA$--independent and analytic in $\e$ if $\ff$ is analytic on
$\TTT^{\ell}$: in that case, as pointed out above, the existence and
analyticity of $\NN_{\e}$ is then implied by Moser's theorem [M1],
Theorem 1. In [M1] it is also pointed out that the latter result is the
``core'' of the proof of the KAM theorem in the analytic case. 

 
However Moser's theorem gives no analyticity result when the
perturbation is non analytic.
So, in this paper we consider \equ(1.1) with $F=0$, or,
equivalently, \equ(1.2) and $\e\ff$ modified into $\e\ff+\NN_\e$, with the
aim of proving the existence {\it and analyticity} of the counterterm
$\NN_\e$ if the perturbation $\ff$ belongs to a class specified below
(see \S 1.3), \ie if it depends {\it non analytically} on the angles in
a special way. The convergence of the perturbative series for the
solution of the equations of motion and for the counterterm is obtained
by taking into account some cancellations which include, besides the ones
necessary to treat the analytic case as in [GM2] (the {\sl infrared
cancellations}), also new cancellations called {\sl ultraviolet
cancellations}).
 
The analyticity (of the solution of the equations of motion and of the
counterterm) in the perturbative parameter $\e$ {\it even though the
interaction $\ff(\aa)$ is non analytic in $\aa$} is a result that we
believe would be difficult to obtain by other tecniques which one could
use to face this problem, like the Moser-Nash smoothing tecnique, [M2].
 
The technical tools for the proof are inherited from [BGGM], where
interaction potentials belonging to the same class of functions are
considered to prove the analyticity in the perturbative parameter of the
KAM invariant tori.
 
\*
 
\0{\bf 1.2.} The paper is organized as follows.
In the remaining part of this section we introduce the notations and
state the result (Theorem 1.4). In \S 2 we reintroduce our diagrammatic
formalism, referring for details to Appendix A1 (or [BGGM], \S 3, and
simply outlining the differences with respect to it).  In \S 3 the so
called {\sl infrared cancellations} are discussed: such cancellations
allow us to solve the problem of small divisors, and they are sufficient
to prove the convergence of the perturbative series in the analytic
case.  Since there are some differences with respect to [GM2], mostly
the use of Siegel-Eliasson's lemma (see Lemma 3.4 below) instead of
Siegel-Bryuno's lemma (see [GM2], Lemma 5.3), we provide a
selfconsistent discussion although the problem is the same as the one
treated in [GM2], \S 8.  In \S 4, we study the {\sl ultraviolet
cancellations} which, together with the infrared ones, make convergent
the perturbative series for the class of interactions introduced in \S
1.3.  In \S 5, we note briefly that Theorem 1.4 for non analytic
interactions does not give really more results than the corresponding
theorem for the analytic case (see [GM2], Theorem 1.4), if applied to
the one-dimensional Schr\"odinger equation with a non analytic quasi
periodic potential; this is a little deceiving, but not quite unexpected
(see comments in \S 5). As a byproduct of the proof, analyticity of
eigenvalues and eigenfunctions in the perturbative parameter is proven
under the assumption that the interaction potential is at least of class
$C^{(p)}$, $p>3\t +1$: this result is new with respect to [Pa].
 
\*
 
Note that, with respect to [BGGM], the discussion of the infared
cancellations appears remarkably less involved (and therefore more
suitable for a first approach to the techniques employed).  On the
contrary the discussion of the ultraviolet cancellations is essentially
unchanged with respect the one in [BGGM], notwithstanding the simplified
expression of the Hamiltonian \equ(1.3) below, and will be here repeated only
for selfconsistence purposes.
 
\*
 
\0{\bf 1.3.} The Hamiltonian is
$$ \HHH = \HHH_0 + \AA\cdot\NN(\e) \; \Eq(1.3) $$
with $ \HHH_0 = \oo_0\cdot\AA +\e \AA\cdot\ff(\aa)$ given by
\equ(1.1), where\\
(1) $\AA\in \RRR^\ell$ and $\aa\in \TTT^\ell$ are
canonically conjugated variables (respectively action
and angle variables), and $\cdot$ denotes
the scalar product;\\
(2) $\oo_0$ is a rotation vector satisfying the {\sl
diophantine condition}
$$ C_0|\oo_0\cdot\nn|>|\nn|^{-\t} \qquad
\forall \nn\in\ZZZ^{\ell}\; , \nn\neq{\V0} \; , \Eq(1.4) $$
for some positive constants $C_0$ and $\t$ (here and henceforth
$|\nn|=\sqrt{\nn\cdot\nn}$, while $\|\nn\|=\sum_{j=1}^\ell |\n_j|$, if
$\nn=(\n_1,\ldots,\n_\ell)$);\\
(3) $\ff$ has the form $\ff=(f_1,\ldots,f_\ell)$, with each $f_j$ of
the class $\hat C^{(p)}(\TTT^\ell)$ introduced in [BGGM], for some $p$:
namely $f_j(\aa) = \sum_{\nn\in\zzz}f_{j\nn}\,e^{i\nn\cdot\aa}$,
$\ff_{\nn}=\ff_{-\nn}$, with $f_{j\V0}=0$ and for $\nn\ne\V0$,
$$ f_{j\nn}=\sum_{n\ge p+\ell}^N\fra{c_n^{(j)}+d_n^{(j)}
(-1)^{||\nn||}}{|\nn|^n}\; , \Eq(1.5) $$
for some $N \ge p+\ell$ and some constants $c_n^{(j)}$, $d_n^{(j)}$; and\\
(4) $\NN(\e)$, called a {\sl counterterm}, and has to
be fixed in order to make the equations of motion soluble
for the model \equ(1.3).
 
For instance we can choose $f_{j\nn}=a_j|\nn|^{-b}$, with $b=p+\ell$ and
$a_j\in\RRR$; then define $\sup_{j=1,\ldots,\ell}|a_j|=a$ and
$\aaa=(a_1,\ldots,a_\ell)$. In the following we shall deal explicitly
with such a function: the proof can be trivially extended to the class
of functions \equ(1.5).
 
\*
 
\0{\bf 1.4.} {\cs Theorem.} {\it
Given the Hamiltonian \equ(1.1), with $\oo_0$ satisfying the diophantine
condition \equ(1.4) and $\ff=(f_1,\ldots,f_\ell)$, with each $f_j\in
\hat C^{(p)}(\TTT^\ell)$, there exist two positive constants $\e_0$ and
$p_0=2+3\t$, and a function $\NN(\e)$ analytic in $\e$ for $|\e|<\e_0$,
such that the equations of motion corresponding to the Hamiltonian
\equ(1.3) admit solutions in $C^{(0)}(\TTT^\ell)$ analytic in $\e$ for
$|\e|<\e_0$.}
 
\*
 
\0{\bf 1.5.} The equations of motion for the Hamiltonian
\equ(1.3) are given by
$$ \eqalign{
{d\a_j\over dt} & = \oo_{0j} +\e f_j(\aa) + N_j(\e)\; ,\cr
{dA_j\over dt} & = - \e\AA\cdot\dpr_{\a_j}\ff(\aa)\; .\cr}\Eq(1.6) $$

We look for solutions of the form
$$ \eqalign{
\aa(t) & = \oo_0 t + \hh(\oo_0t) \; , \qquad \hh(\pps) =
\sum_{k=1}^{\io} \sum_{\nn\in\zzz}
\hh^{(k)}_\nn \, e^{i\nn\cdot\pps},\e^k \; , \cr
\AA(t) & = \AA_0 + \HH(\oo_0t) \; , \qquad \HH(\pps) =
\sum_{k=1}^{\io} \sum_{\nn\in\zzz}
\HH^{(k)}_\nn \, e^{i\nn\cdot\pps}\,\e^k \; , \cr} \Eq(1.7) $$

\0with $\V h$ {\it odd} and $\V H$ {\it even} in $\V\ps$ so that the
equations for $\hh$ and $\HH$ become
$$ \eqalign{
(\oo_0\cdot\dpr_\pps)\,h_j(\pps) & = \e f_j(\pps+\hh(\pps)) +
N_j \; , \cr
(\oo_0\cdot\dpr_\pps)\,H_j(\pps) & = -\e [\AA_0
+\HH(\pps)]\cdot\dpr_{\a_j}\ff(\pps+\hh(\pps)) \; , \cr} \Eq(1.8) $$
where $\dpr_{\aa}$ denotes derivative with respect to the argument, and
$\NN$ has to be so chosen that the right hand side of the first equation
in \equ(1.8) has vanishing average.
 
We see from \equ(1.8) that the equation for $\hh$ is closed, so that, as
far as one is interested only in the function $\hh$, one can confine
ourself to study only the first equation in \equ(1.8). This is the
equation that one has to solve to linearize the flow generated by
$d\aa/dt=\oo_0+\e\ff(\aa)+\NN_\e$: hence it is not surprising that the
equation for $\V H$ can be easily solved once $\hh$ is known: see
\S 2.3 below.
 
Note that, since $\V f$ is supposed even, then we expect that $\hh$ is
odd and $\HH$ is even: hence while the equation for $\V H$ does not seem
to hit any obvious compatibility problems we see that the equation for
$\V h$ does, {\it unless $N_j$ is suitably chosen}. In fact the function
$\e\ff(\pps+\hh(\pps))$, being even, has no \ap reason to have a vanishing
integral over $\pps$ (as it should, being equal to $(\oo_0\cdot\V
\dpr_\pps)\hh(\pps)$).
 
 
\vskip1.truecm

\centerline{\titolo 2. Formal solutions and graph representation}
\*\numsec=2\numfor=1
 
\0{\bf 2.1.} We study now the equations \equ(1.8) with
$f_{j\nn}=a_j|\nn|^{-b}$ replaced with $f_{j\nn}\,e^{-\k|\nn|}$.
The parameter $\k$ is taken $\k>0$, and, after
computing the coefficients $\hh^{(k)}_\nn$ in \equ(1.7),
which will depend on $\k$, one will perform the limit $\k\to0$
({\sl Abel's summation}).
 
The formal solubility of \equ(1.8) with $f_{j\nn}$ replaced
with $f_{j\nn}\,e^{-\k|\nn|}$ follows from [GM2], \S 8.1,
where more general interaction potentials are considered.
 
One has $h^{(k)}_{j\V0}=H^{(k)}_{j\V0}=0$ $\forall k\ge1$,
while, when $\nn\neq\V0$, for $k=1$,
$$ h^{(1)}_{j\nn} = { f_{j\nn_0} \over i\oo_0\cdot\nn} \; , \qquad
H^{(1)}_{j\nn} = - { i\n_{0j} \over i\oo_0\cdot\nn}
\Big( \AA_0 \cdot \ff_{\nn_0} \Big) \; , \Eq(2.1) $$
and, for $k\ge 2$,
$$ \eqalignno{
h^{(k)}_{j\nn} = &
{1\over i\oo_0\cdot\nn} \sum_{p>0} {1\over p!}
\sum_{\nn_0+\nn_1+\ldots+\nn_p=\nn} f_{j\nn_0}
\sum_{k_1 + \ldots + k_p = k-1}
\prod_{s=1}^p \Big(i\nn_0\cdot\hh_{\nn_s}^{(k_s)}\Big) \; , \cr
H^{(k)}_{j\nn} = &
- {1\over i\oo_0\cdot\nn} \sum_{p>0} {1\over p!}
\sum_{\tilde \nn + \nn_0 + \nn_1 + \ldots + \nn_p = \nn} (i\n_{0j})
\; \cdot \cr & \hskip1.truecm \cdot
\sum_{\tilde k+ k_1+\ldots + k_p = k-1}
\Big( \HH_{\tilde\nn}^{(\tilde k)} \cdot \ff_{\nn_0} \Big)
\prod_{s=1}^p \Big(i\nn_0\cdot\hh_{\nn_s}^{(k_s)}\Big) \cr
& - {1\over i\oo_0\cdot\nn} \sum_{p>0} {1\over p!}
\sum_{\nn_0+\nn_1+\ldots+\nn_p=\nn} (i\n_{0j})
\; \cdot&\eq(2.2) \cr & \hskip1.truecm \cdot
\sum_{k_1 + \ldots + k_p = k-1}
\Big( \AA_0 \cdot \ff_{\nn_0} \Big)
\prod_{s=1}^p \Big(i\nn_0\cdot\hh_{\nn_s}^{(k_s)}\Big)
\; , \cr} $$
provided
$N_j(\e)=\sum_{k=1}^{\io}N_j^{(k)}\,\e^k$, with
$N_j^{(k)}$ defined by
$N_j^{(1)}= - f_{j\nn_0}$ and, for $k\ge 2$,
$$ N^{(k)}_{j} =
- \sum_{p>0} {1\over p!}
\sum_{\nn_0+\nn_1+\ldots+\nn_p=\V0} f_{j\nn_0}
\sum_{k_1+\ldots+k_p= k-1}
\prod_{s=1}^p \Big(i\nn_0\cdot\hh_{\nn_s}^{(k_s)}\Big) \; . \Eq(2.3) $$

The equality \equ(2.3) assures the formal solubility of \equ(1.8).
Note that, as $\ff$ is even, then $\hh$ is odd and $\HH$ is even.
 
If $\ff$ is analytic ($\k>0$) the convergence of the series defining the
functions $\hh$ and $\HH$ is a corollary of [GM2], Theorem 1.4, but the
convergence radius is not uniform in $\k$ (it shrinks to zero when
$\k\to0$).  The aim of the present paper is to show that, if $\ff$
belongs to the class of functions $\hat C^{(p)}(\TTT^{\ell})$, then
there are cancellation mechanisms assuring the convergence of the series
and so the analyticity in $\e$ of the solutions of the equations of
motion.
 
\*
 
\0{\bf 2.2.} We shall use a representation of \equ(2.2) in terms of
{\it Feynman graphs} following the rules in [BGGM], \S 3: the reader not
familiar with [BGGM] can find in Appendix A1 below a brief but
selfconsistent description of the graphs. See [BGM] for the terminology
motivation.  The only difference will be that that the ``value'' of a
graph $\th$ is now given by
$$ \Val (\th) = \prod_{v<r} { (i\nn_{v'}\cdot\ff_{\nn_v})
\over i\oo_0\cdot\nn_{\l_v} } \; , \Eq(2.4)$$
where\\
(1) $v'$ is the node immediately following $v$ in $\th$, and
$\l_v=v'v$ is the line emerging from $v$ and entering $v'$;\\
(2) $r$ is the {\sl root} of the graph, and
$i\nn_r$ denotes the unit vector in the $j$th direction,
$i\nn_r=\ee_j$, $j=1,\ldots,\ell$;\\
(3) $\nn_v$ is the {\sl external momentum}
associated to the {\sl node} $v$,
$\nn_{\l_v} = \sum_{w\le v} \nn_w$ is the {\sl momentum} flowing
through the line $\l_v$, and $\nn(\th)$ is the momentum
flowing through the line entering the root.
 
It can be convenient to introduce the notations
$$ g_{\l}\={1\over \oo_0\cdot\nn_{\l}} \; , \qquad
D(\th) = \prod_{\l\in\th} g_{\l} \; , \Eq(2.5) $$
so that \equ(2.3) becomes
$$ \Val(\th) = D(\th) \prod_{v<r} \left(\nn_{v'}\cdot\ff_{\nn_v}
\right)\; ;\Eq(2.6) $$
$g_\l$ is called the propagator of the line $\l$.
Let us denote by $\TT (k,\nn)$ the set of non equivalent labeled
graphs of order $k$ with $\nn(\th)=\nn$ and $i\nn_r=\ee_j$; then
$$ h^{(k)}_{j\nn} = {1\over k!}
\sum_{\th \in \TT(k,\nn)} \Val(\th) =
{1\over k!} \sum_{\th^0\in\TT^0(k)} W(\th^0,\nn) \; , \Eq(2.7) $$
where $\th=(\th^0,\{\nn_\l\})$, if $\th$ is a labeled graph,
while $\th^0$ is a graph bearing no external momentum labels,
$\TT^0(k)$ is the set of such graphs of order $k$, and
$$ W(\th^0,\nn) \= \sum_{ \{\nn_x\} :\,
\nn(\th)=\nn} \Val(\th) \; . \Eq(2.8) $$

\*
 
\0{\bf 2.3.} The function $H_{j\nn}^{(k)}$ can be expressed in terms
of the same graphs as in \S 2.2, but the value associated to
a graph $\th$ is no more given by \equ(2.3). On the contrary
one define
$$ \eqalign{
\hbox{Val}^*(\th) = & D(\th) \, (-i\n_{0j})
\sum_{\tilde v \in \th}
\Big[ \prod_{v\notin \CC(v_0,\tilde v) }
(\nn_{v'}\cdot\ff_{\nn_v}) \Big] \cdot \cr
& \cdot
\Big[ \prod_{v\in \CC(v_0,\tilde v) \setminus \tilde v}
(-\ff_{\nn_v}\cdot\nn_{v'{'}}) \Big] (\ff_{\nn_{\tilde v}}\cdot
\AA_0) \; , \cr} \Eq(2.9) $$
where\\
(1) $v_0$ is the highest node in $\th$, \ie $v_0'=r$,\\
(2) $\CC(v_0,\tilde v)$ is the collection of vertices
crossed by the connected path of branches
in $\th$ linking the node $v_0$ with a node $\tilde v\le v_0$,
with $\CC(v_0,v_0)=v_0$,\\
(3) $v'{'}$ is the node on $\CC(v_0,\tilde v)$ immediately
preceding $v$.
 
With the above definition \equ(2.8), one has
$$ H_{j\nn}^{(k)} = {1\over k!}
\sum_{\th\in\TT(k,\nn)} \hbox{Val}^*(\th) \; , \Eq(2.10) $$
where $\TT(k,\nn)$ is defined as after \equ(2.6).

\vskip1.truecm

\centerline{\titolo 3. Infrared cancellations}
\numsec=3\numfor=1
\*
\0The cancellations discussed in this section are cancellations that are
sufficient to treat the problem in the analytic case. Hence they are
not really characteristic of the problems that we address in this paper.
Nevertheless they must be taken into account and their compatibility
with the cancellations that are typical of the differentiable problem
will have to be, eventually, discussed.
\*
 
\0{\bf 3.1.} Let us define $\chi(x)$ as the characteristic function
of the set $\{x\in \RRR : |x|\in[1/2,1)\}$, and $\chi_1(x)$ as the
characteristic function of the set $\{x\in\RRR, |x|>1\}$.  Then each
propagator in \equ(2.4) can be decomposed as
$$ \eqalign{
g_{\l} & = {\ch_1(\oo_0\cdot\nn_{\l}) \over \oo_0\cdot\nn_{\l}}
+ \sum_{n=-\io}^0 { \ch(2^{-n} \oo_0\cdot\nn_{\l})
\over \oo_0\cdot\nn_{\l} }
= \sum_{n=-\io}^1 g_{\l}^{(n)} \; ,\cr } \Eq(3.1)$$
and, inserting the above decompositions in the definition of the value
of a graph \equ(2.3), we see that the value of each graph
is naturally decomposed into various addends. We can identify the addends
simply by attaching to each line $\l$ a scale label $n_\l\le1$.
 
\*
 
\0{\bf 3.2.} {\cs Definition (Cluster)}.
{\it Given a graph $\th$, a {\sl cluster} of scale $n\le1$
is a maximal set of nodes connected by lines of scale $\ge n$.
A line $\l$ which connects nodes both inside a cluster
$T$ is said to be {\sl internal} to the cluster: $\l \in T$;
the lines which connect a node inside with a node outside the cluster are
called {\sl external} to the cluster;
if a line $\l$ is internal or external to a cluster $T$,
we say that $\l$ intersect $T$: $\l\cap T \neq \emptyset$.
A line is {\sl outside} the cluster $T$ if it is neither internal nor
external to it.}
 
\*
 
The nodes of a cluster $V$ of scale $n_V$ may be linked to other nodes
by lines of lower scale.  Such lines are called {\sl incoming}
if they point at a node in the cluster or {\sl outgoing} otherwise;
there may be several incoming lines (or zero) but at most
one outgoing line, because of the tree structure of the graphs.
 
\*
 
\0{\bf 3.3.} {\cs Definition (Resonance).}
{\it We define a {\sl resonance} a cluster $V$ such that:\\
(1) there is only one incoming line $\l_V$ and one outgoing line
$\l'_V$ and $|\nn_{\l_V}|=|\nn_{\l_V'}|$;\\
(2) if $n_V$ is the scale of the cluster and $n_{\l_V}$ is the scale
of the line $\l_V$, one has $n_V\ge n_{\l_V}+3$.\\
If $\nn_{\l_V}=\nn_{\l_V'}$, the resonance is called a {\sl real
resonance}, if $\nn_{\l_V}=-\nn_{\l_V'}$, it is called
a {\sl virtual resonance}.}
 
\*
 
Then the following result holds (see [S,E,BGGM]):
 
\*
 
\0{\bf 3.4.} {\cs Lemma (Siegel-Eliasson's bound).}
{\it If we consider only graphs with no real
resonances, then
$$ \prod_{\l\in\th} \fra1{|\oo_0\cdot\nn_{\l}|}
\le C^{k} \,\fra{ \prod_{v\in\th} |\nn_v|^{{\h\over2}\t}}
{(\sum_{v\in\th}|\nn_v|)^{\t}} \; , \Eq(3.2) $$
for some positive constant $C$ and $\h=6$.}
 
\*
 
\0{\bf 3.5.}
Consider a graph $\th$ and call $\hat\th$ the graph obtained by deleting
the infrared scale labels $\{n_\l\}$ and $\th^0$ the graph obtained by
deleting the scale and external momentum labels:
$\th=(\hat\th,\{n_\l\})$, or $(\th^0,\{n_\l\},\{\nn_x\})$.
 
Suppose that the set of scales $\{n_\l\}$ is consistent with the
existence of a fixed family $\V V_1$ of {\sl maximal real resonances}:
\ie of real resonances not contained in any larger real resonance.
If $V\in \V V_1$ we call $\l_V=v_V^b v_V^1$ the line incoming
into the real resonance and $n_{\l_V}$ its scale;
likewise $\l_V'=v_V^0 v_V^a$ is the outgoing line,
($v_V^a,v_V^b\in V$ while $v_V^0,v_V^1$ are out of it).
 
We consider the graph values at fixed set of scales for the lines not in
any $V\in \V V_1$, and we say that such a set of scales is ``compatible"
with $\V V_1$, denoting this property by $\{n_\l\}\&\V V_1$.
 
We introduce the {\sl momentum flowing on $\l_v\in V$ intrinsic to
the cluster $V$} as $\nn_{\l_v}^0=\sum_{v\ge w\in V}\nn_w$,
and define the {\sl resonance path} $Q_V$
as the totally ordered path of lines joining the
line incoming into the real resonance $V$ and the outgoing line
and {\it not} including the latter two lines. Then
$$ \eqalignno{
\sum_{\{n_\l\}} \Val & (\hat\th, \{n_\l\})=
\sum_{\V V_1}\sum_{\{n_\l\}\& \V V_1}
\cdot \Big[ \prod_{\l \cap \V V_1  = \emptyset \atop \l=xy}
(\nn_x\cdot\ff_{\nn_y})\,{ \ch(2^{-n_\l}\oo_0\cdot\nn_\l) \over
\oo_0\cdot\nn_\l } \Big] \cdot & \eq(3.3) \cr
& \cdot \prod_{V\in\V V_1} \Big\{
\Big[ (\nn_{v_0^V}\cdot\ff_{\nn_{v_V^a}})
\, (\nn_{v_b^V}\cdot\ff_{\nn_{v_V^1}} )\,
{ \ch(2^{-n_\l}\oo_0\cdot\nn_{\l_V}) \over
(\oo_0\cdot\nn_{\l_V})^2} \Big]
\VVV(\oo_0\cdot\nn_{\l_V}|V,\{n_\l\}_{\l\in V})\Big\} \;,\cr}$$
where the {\sl resonance value} $\VVV$ is defined by
$$ \VVV (\z| V, \{n_\l\}_{\l\in V})=
\prod_{\l\in V \atop \l=xy} (\nn_x\cdot\ff_{\nn_y})\,
{ \ch(2^{-n_\l}(\oo\cdot\nn^0_\l+\s_\l \z)) \over
\oo\cdot\nn^0_\l+\s_\l\z } \; , \Eq(3.4) $$
with $\s_\l=1$ if $\l$ is on the resonance path $Q_V$,
($\l\in Q_V$), else $\s_\l=0$.
 
\*
 
Let $\ch^{(n,n')}(x)$ denote the characteristic function of the set
$|x|\in [2^{n-1},2^{n'})$; then \equ(3.3) becomes
$$ \sum_{ \{\nn_{\l} \} } \Val(\hat\th,\{\nn_\l\}) =
\sum_{\V V_1} \sum_{ \{\nn_\l\}_{\l\notin \V V_1} \& \V V_1}
X_1 (\hat\th,\{\nn_\l\}) \; , $$
where
$$ \eqalignno{
X_1 (\hat\th, & \{n_\l\})=
\sum_{\V V_1}\sum_{\{n_\l\}_{\l\in \V V_1} \& \V V_1}
\cdot \Big[ \prod_{\l \cap \V V_1  = \emptyset \atop \l=xy}
(\nn_x\cdot\ff_{\nn_y})\,{ \ch(2^{-n_\l}\oo_0\cdot\nn_\l) \over
\oo_0\cdot\nn_\l } \Big] \cdot & \eq(3.5) \cr
& \cdot \prod_{V\in\V V_1} \Big\{
\Big[ ( \nn_{v_V^0}\cdot\ff_{\nn_{v_V^a}})
\,(\nn_{v_V^b}\cdot\ff_{\nn_{v_V^1} })\,
{\ch(2^{-n_\l}\oo_0\cdot\nn_{\l_V}) \over (\oo_0\cdot\nn_{\l_V})^2} \Big]
\VVV'(\oo\cdot\nn_{\l_V}|V,\{n_\l\}_{\l\in V})\Big\}\;, \cr} $$
with
$$ \VVV'(\z| V, \{n_\l\}_{\l\in V}) =
\prod_{\l\in V \atop \l=xy}
(\nn_x\cdot\ff_{\nn_y})\,
{\ch^{(n_{\l_V}+3,+\io)}(\oo\cdot\nn^0_\l+\s_\l\z)
\over (\oo\cdot\nn^0_\l+\s_\l\z)^2 } \; , \Eq(3.6)$$
which can be rewritten, by using the Lagrange's interpolation
formula,
$$ \VVV'(\z| V, \{n_\l\}_{\l\in V}) =
\VVV'(0| V, \{n_\l\}_{\l\in V}) +
\VVV_1'(\z| V, \{n_\l\}_{\l\in V}) \; , $$
where
$$ \eqalign{
\VVV_1'(\z| V, \{n_\l\}_{\l\in V}) & =
\Big( \prod_{\l\in V/Q_V \atop \l=xy}
(\nn_x\cdot\ff_{\nn_y})\,
{ \ch^{(n_{\l_V}+3,+\io)}(\oo_0\cdot\nn^0_\l) \over \oo_0\cdot\nn^0_\l }
\Big) \cdot
\cr & \cdot
\z \ig_0^1 dt_V \,
\fra{\dpr}{\dpr t_V}
\Big[\prod_{\l\in Q_V\atop \l=xy} (\nn_x\cdot\ff_{\nn_y})\,
\fra{\ch^{(n_{\l_V}+3,+\io)}(\oo_0\cdot\nn^0_\l+t_V
\z)}{\oo_0\cdot\nn^0_\l+t_V\z}\Big]
\; . \cr} \Eq(3.7)$$

We can consider the value $X_1$ with the real resonance
value corresponding to $V\in\V V_1$ simply replaced by the
expression defined in \equ(3.7); this
follows from the following result.
 
\*
 
\0{\bf 3.6} {\cs Lemma.} {\it When all graphs
in \equ(3.3), with the real resonance value
$\VVV'(\z| V, \{n_\l\}_{\l\in V})$ replaced
with $\VVV'(0| V, \{n_\l\}_{\l\in V})$, are summed together,
they give a vanishing contribution.}
 
\*
 
The proof is in Appendix A2.
 
\*
 
\0{\bf 3.7.}
We can perform explicitly the derivative in \equ(3.7):
we obtain (see also Remark after (5.6) in [BGGM])
$$ \eqalign{
X_1(\hat\th,\{\nn_\l\}) & =
\Big( \prod_{\l\in\hat\th\atop\l=xy}\nn_x\cdot\ff_{\nn_y} \Big)
\cdot \Big( \prod_{\l\in \hat \th/\V V_1}\fra{\ch(2^{-n_\l}
(\oo_0\cdot\nn_\l))} {\oo_0\cdot\nn_\l} \Big) \cdot
\cr
& \cdot \sum_{\l_V^0\in Q_V}
\sum_{z=0}^1\ig_0^1dt_V\,
p(\l_V^0,z,t_V) \, \cdot \cr
& \cdot \prod_{V\in \V V_1} \Big[ \prod_{\l\in V}
\fra{\ch^{(n_{\l_V}+3,+\io)}
(\oo_0\cdot\nn_\l(t_V))} {\oo_0\cdot\nn_\l(t_V)} \Big]
\; , \cr} \Eq(3.8) $$
where, for $\l\in V$, we adopt the notation
$\nn_{\l}(t_V)=\nn_{\l}^0+t_V\nn_{\l_V}$, and
$$ p(\l_V^0,z,t_V) =
\cases{ - \fra{\oo_0\cdot\nn_{\l_V}}{\oo_0\cdot\nn_{\l_V^0}(t_V)} \; ,
\qquad & if $\ z=1 \; $, \cr
\sum_{t^*_V}\d(t_V-t^*_V) \; , \qquad & if $z=0 \; $, \cr} \Eq(3.9) $$
where $t^*_V$ are the solutions to the equation
$|\oo_0\cdot\nn_{\l_V^0}(t_V)|=2^{n_{\l_V}+2}$, if any
(there are at most 2 solutions).
 
We then redecompose in \equ(3.8) the characteristic functions
of the lines inside the real resonances into individual scales
from $n_{\l_V}+3$ up, so that \equ(3.5) becomes
$$ \eqalignno{
X_1(\hat\th,\{\nn_\l\}) & =
\sum_{\{n_\l\}_{\l\in \V V_1}\& \V V_1}
\Big[ \Big( \prod_{V\in \V V_1} \ig_0^1 dt_V  \Big)\cdot
\prod_{\l\in\hat\th,\, \l=xy \atop \l\notin \{\l_V'\}_{V\in \V V_1}}
\Big( { \ch(2^{-n_\l}\oo_0\cdot\nn_{\l}(\V t)) \over
\oo_0\cdot\nn_{\l}(\V t) } \Big) \cdot \cr
& \cdot \Big( \prod_{\l \in\hat\th
\atop \l=xy} \nn_x\cdot\ff_{\nn_y} \Big)
\cdot \Big( \prod_{V\in \V V_1} \sum_{\l_V^0\in Q_V}
\sum_{z_V=0}^{1}
{d^{z_V}_{\l_V^0} \over \oo_0\cdot\nn_{\l_V^0}(\V t) }
\Big) \Big]  \; , & \eq(3.10) \cr}$$
where $\V t=\{t_V\}_{V\in \V V_1}$ and we set
$\nn_{\l}(\V t) = \nn_{\l}^0+t_V\nn_{\l_V}$ if $\l\in Q_V$, and
$\nn_{\l}(\V t)= \nn_{\l}^0\=\nn_{\l}$ if $\l \notin \cup_{V\in V_1}
Q_V$, and
$$ d^1_{\l_V^0} = - 1 \; , \qquad
d^0_{\l_V^0} =
\tst{ \fra{\oo\cdot\nn_{\l_V^0}(\V t)}{\oo\cdot\nn_{\l_V}}}\,
\sum_{t^*_V}\d(t_V-t^*_V) \; , \Eq(3.11)$$
where $t_V^*$ is defined as in \equ(3.9).
 
Each addend in \equ(3.10), with fixed $\th=(\hat\th,\{n_\l\})$
and $\{\l_V^0, z_V,t_V\}_{V\in \V V_1}$, is said to be
{\sl superficially renormalized} on the real
resonances $\V V_1$, on the line $\l_V^0$ and on the
choices $z_V$.
 
\*
 
\0{\bf 3.8.} {\cs Remark.}
Note that the case $z_V=0$ is special as it forces
$n_{\l_V^0}=n_{\l_V}+3$, so that the ratio in the
definition \equ(3.11) of $d^0_{\l_V^0}$ is bounded above by $2^4$.
 
\*
 
\0{\bf 3.9.} Having dealt with the maximal real resonances ({\sl first
generation} real resonances) we perform again the same operations: \ie
fixed $\hat\th$, $\V V_1,\{\l_V^0,$ $z_V, t_V\}_{V\in \V V_1}$ and the
scales $\{n_\l\}$ for $\l\not\in \cup_{V\in \V V_1}V$, we identify the
{\sl second generation} real resonances as the maximal real resonances
inside each $V\in \V V_1$; call $\V V_2$ the set of the real resonances
of the first {\it and} second generations and proceed in a similar way
to ``renormalize" superficially the newly considered real resonances
$W\in \V V_2/\V V_1$.
 
This means that we fix the scale labels of the lines outside the two
generations of real resonances, and sum over the other scale labels
$\{n_\l\}$ consistent with the elements of $\V V_2$ being the first and
second generation resonances.
 
We obtain that the product in \equ(3.10)
of the terms coming from the lines $\l\in W\in \V V_2$, $W\subset
V\in\V V_1$ can be written in a form very close to \equ(3.6),
with the difference that the momenta flowing through the
lines $\l\in Q_W\cap Q_V$ are $\nn_\l(\V t)= \nn^0_\l+t_W\,
(\nn^0_{\l_W}$ $+$ $t_V\nn_{\l_V})$, and $n_{\l_V}+3$ is replaced
by $n_{\l_W}+3$ in the characteristic functions.
 
We proceed to perform a Taylor expansion as above, in the variables $\z_W=
\oo_0\cdot(\nn^0_{\l_W}+t_V\nn_{\l_V})$ if $\l_W\in Q_V$ or
$\z_W=\oo_0\cdot\nn^0_{\l_W}$ otherwise.
However this time we modify the renormalization procedure:
if $W$ contains the line $\l_V^0$ we do nothing, while
if $\l_V^0\notin W$ we write the first order remainder as above.
 
We then perform the derivatives with respect to the new interpolation
parameters $t_W$, generated by the expression of the Taylor
series remainders, hence redevelop the
characteristic functions and rearrange, along the lines that generated
\equ(3.10), the various terms to simplify the notation and to get an
expression very similar to \equ(3.10), for a quantity that we
could call $X_2$. Note that the order zero term of each Taylor
expansion can be neglected, because of Lemma 3.5, which still hold
if $V$ is not a maximal resonance, but is contained in another
resonance.
The latter $X_2$ can subsequently be used, in the same way as the $X_1$
was already used to start the second renormalizations, for the
superficial renormalization of the third generation of real resonances.
Then we iterate step by step the procedure until there are no more
real resonances inside the maximal real resonances found at the last step
performed and all the $n_\l$ have been fixed.
 
\*
 
To write down the final expression we need some notations.\\
(1) Let us call $\V V$ the collection of all real resonances
selected along the iterative procedure.
For each $V\in \V V$ choose a line $\l_V$ with the
{\sl compatibility condition} that if $V\subset Z\in\V V$,
$\l_Z^0\in V$ implies $\l_V^0=\l_Z^0$.\\
(2) Then if $\l_Z^0\notin V$ we say that the line $\l_V^0$
is {\sl new} and that the real resonance $V$ is {\sl new}, and define
$$ \eqalign{
& \p_V(dt_V) = dt_V \; , \cr
& d^1_{\l_V^0} = - 1 \; , \qquad
d^0_{\l_V^0} =
\tst{ \fra{\oo_0\cdot\nn_{\l_V^0}(\V t)}{\oo_0\cdot\nn_{\l_V}} }\,
\sum_{t^*_V} \d(t_V-t^*_V) \; , \cr} \Eq(3.12)$$
where $t_V^*$ are the solutions (at most 2) of the
equation $|\oo_0\cdot\nn_V^0(\V t)| = 2^{n_{\l_V}}+2$ for $t_V$,
and the interpolated momenta $\nn_{\l}(\V t)$ are defined as
$$ \nn_{\l}(\V t) = \cases{
\nn_{\l} \; , & if $\l$ is not contained in any resonance paths, \cr
\nn_{\l} \prod_{V \; :  \l \in Q_V} t_V \; , & otherwise $\;$. \cr}
\Eq(3.13) $$
\\(3) If $\l_Z^0\in V$ (so that $\l_Z^0=\l_V^0$) we say
that $\l_V^0$ and $V$ are {\sl old}, and define
$$ \eqalign{
& \p_V(dt_V) = \d (t_V-1)\,dt_V \; , \cr
& d^1_{\l_V^0} = 1 \; , \qquad
d^0_{\l_V^0} = 0 \; . \cr} \Eq(3.14) $$
\\(4) Define
$$ P_0(\th)=\prod_{\l\not\in \cup_V \l_V'}\fra{\ch
(2^{-n_\l}\oo_0\cdot\nn_\l(\V t))}{\oo_0\cdot\nn_\l(\V t)},\qquad
N(\th)=\prod_{\l\in\hat\th\atop\l=xy}\nn_x\cdot\ff_{\nn_y}
\; , \Eq(3.15)$$
and denote by $\L$ the function $V\to\{\l_V^0,z_V\}$.\\
(5) Define
$$\RR \Val(\th)= N(\th)\,\RR D(\th) \; , \Eq(3.16)$$
where
$$ \RR D(\th) = \sum_\L \prod_{V\in \V V} \ig_0^1 \p(dt_V)\,
P_0(\th) \prod_{V\in\V V} {d^{z_V}_{\l_V^0} \over
(\oo_0\cdot\nn_{\l_V^0} (\V t))^* }  \; , \Eq(3.17) $$
and the ${}^*$ means that
$(\oo_0\cdot\nn_{\l_V^0}(\V t))^*=\oo_0\cdot\nn_{\l_V^0}(\V t)$
if the resonance is new, and
$(\oo_0\cdot\nn_{\l_V^0}(\V t))^*=\oo\cdot\nn_{\l_V}$
if the resonance is old.
 
\*
 
\0{\bf 3.10.} {\cs Remark.} By Definition 3.3,
we have $|\oo_0\cdot\nn_{\l}(\V t)|\ge \fra23 |\oo_0\cdot\nn_\l^0|$,
uniformly in $\V t$.
 
\*
 
\0{\bf 3.11.}
Then, from the iterative procedure and with the just introduced
notations, we obtain
$$ \sum_\th \Val(\th) = \sum_\th \RR \Val (\th) \; , \Eq(3.18) $$
as all terms discarded in each Taylor expansion add to
zero when summed together (as a corollary of Lemma 3.6).
 
The number of terms thus generated is, at fixed $\V V$, bounded by the
product over $V\in \V V$ of $2$ times the number of pairs that are in
$V/\cup_{W\subset V, W\in\V V} W$ and therefore it is bounded by $2^k
\prod_V k(V)^2$ if $k(V)$ is the number of nodes in $V$ which are not
in real resonances inside $V$.  Hence this number is $\le (2^4)^k$.
The number of families of real resonances in $\hat\th$
({\it hence at fixed $\{\nn_x\}$}) is also bounded by $2^k$.
 
\*
 
\0{\bf 3.12.}
After applying the $\RR$ operations, we see that the contribution to
the new ``re\-nor\-malized value" from the divisors in \equ(3.10) will
be bounded by the same product appearing in the non renormalized values
of the graphs {\it deprived of the divisors due to the lines exiting
re\-so\-nan\-ces} times a factor
$$ \eqalign{
2^{4k}\prod_{V \subset \th}
\fra{1}{\min_{\l\in V_0}
|\oo_0\cdot\nn_{\l}(\V t)|}
& \le C_1^k \prod_{V \subset \th}
\fra{1}{\min_{\l\in V_0} |\oo_0\cdot\nn_{\l}^0|} \le \cr
& \le C_1^k \prod_{V\subset \th}\Big[\sum_{v\in V_0} |\nn_v| \Big]^{\t}
, \cr} \Eq(3.19) $$
where the factor $2^{4k}$ arises from Remark 3.8 and
$C_1=3/2$ from Remark 3.10, and
$V_0$ is the set of nodes inside the real resonance $V$
not contained in the real resonances internal to $V$.
 
Then, we can identify the real resonances $V\in \V V$ of different
generations.  The set $\V V_j$ of real resonances of the $j$th
generation, $j\ge1$, just consists of the real resonances which are
contained in $(j-1)$th generation real resonances (of lower scale) but
not in any $(j+1)$th generation real resonances.
 
If $V$ is a real resonance in $\V V_j$ with entering line $v_V^b v_V^1$
and outgoing line $v_V^a v_V^1$ with momentum $\nn_{\l_V}$ we can
construct a ``$V$-contracted graph" by replacing the cluster $V$
together with the incoming and outgoing lines by the single line
$v_V^0v_V^1$: \ie by deleting the resonance $V$ and replacing it by a
line.  We can also construct the ``$V$-cut graphs'' by deleting
everything but the lines of the resonance $V$ and its entering and
outgoing lines and, furthermore, by deleting the outgoing line as well
as the node $v_a^V$ and attributing to the node $v_V^1$ an external
momentum equal to the momentum flowing into the entering line in the
original graph $\th$: thus we get $p_{v_V^a}$ disconnected graphs.
 
We repeat the above two operations until we are left only with graphs
$\th_i$, $i=1,2,\ldots$ without real resonances: by construction
the product $\prod_{\l\in\th} |\oo_0\cdot\nn_\l(\V t)|^{-1}$
is the same as the $\prod_i\prod_{\l\in\th_i}
|\oo_0\cdot\nn_\l(\V t)|^{-1}$.
 
Then we imagine to delete as well the lines of the various $\th_j$
which were generated by the old entering lines (not all $\th_i$ contain
such lines, but some do) and we call $\th^0_i$ the graphs so obtained.
By doing so we change the momenta flowing into the lines of the graphs
$\th_i$ by an amount which is either $\V0$ or the old momentum
$\nn_{\l_V}(\V t)$ entering a real resonance $V$,
and from Remark 3.10, we have
$$ \prod_{\l\in\hat\th} |\oo_0\cdot\nn_\l(\V t)|^{-1}
\le C_1^k \prod_i \prod_{\l\in\th^0_i} |\oo_0\cdot\nn^0_\l|^{-1}
\prod_{\l\in\{\l_V'\}_{V\in \V V}} |\oo_0\cdot\nn_\l(\V t)|^{-1}
\; , \Eq(3.20) $$
where the last product can be bounded by \equ(3.18),
while Lemma 3.4 gives
$$ \prod_i \prod_{\l\in\th^0_i} |\oo_0\cdot\nn^0_\l|^{-1}
\le C^{k}
\prod_i \fra{\prod_{v\in\th_i^0} |\nn_v|^{\fra{\h}{2}\t}}
{ (\sum_{v\in\th^0_i}|\nn_v|)^{\t}}
,\Eq(3.21) $$
with $\h=6$. Then \equ(3.17)$\div$\equ(3.21) imply
$$ \Big| \RR D(\th) \Big| \le C^{2k}C^{2k}_1 \prod_{v\in\th}
|\nn_v|^{ {\h\over2}\t} \; , \qquad \h=6 \; . \Eq(3.22) $$.

\vskip1.truecm

\centerline{\titolo 4. Ultraviolet cancellations}
\numsec=4\numfor=1
\*
 
\0The ultraviolet cancellations are characteristic of the linearization
problems relative to \equ(1.1), \equ(1.2). They are quite different from
the infrared ones discussed in \S3 and the main technical problem,
besides their identification, is their compatibility with the infrared
cancellations. Exhibiting the two cancellations may not be possible
simultaneously in the sense that the first cancellations may require
grouping graphs in classes that are completely different from the
groupings that are necessary to exhibit the second cancellations. If
this happens one says that the cancellations are not independent and it
is clear that one runs into serious problems.
 
Hence the analysis that follows will be mostly devoted to showing that,
besides an obvious incompatibility that can be explicitly resolved,
the two cancellations are in fact independent.

\*
 
\0{\bf 4.1.} Given a graph $\th$, we can define the {\it
scale} $h_v$ of the node $v$ to be the integer $h_v\ge1$ such that
$2^{h_v-1}\le|\nn_v|<2^{h_v}$.  We say that the labels $\{\nn_x\}$ and
$\{h_x\}$ are {\it compatible} if $|\nn_v|\in[2^{h_v-1},2^{h_v})$ for all
$v\in\th$.  The compatibility relationship between $\{\nn_x\}$ and
$\{h_x\}$ will be denoted $\{\nn_x\}{\,\rm comp\,}\{h_x\}$.  \*
 \*
 
Then we can write
$$ \sum_{\th} \RR\hbox{Val}(\th) = \sum_{\th^0} \sum_{\{\nn_x\}}
\RR\hbox{Val}(\th^0,\{\nn_x\}) = \sum_{\th^0} \sum_{\{h_x\}}
\sum_{\{\nn_x\}{\rm\,comp\,}\{h_x\}} \RR\hbox{Val}(\th^0,\{\nn_x\}) \; .
\Eq(4.1) $$

Set $\QQ=\cup_{V\in\V V}Q_V$, where $\V V$ is the set of all
resonances of $\th$ and $Q_V$ is the resonance path of the
resonance $V$ (see \S 3.5).
Define $\BB_{v}$ the subset of the nodes $w$
among the $p_{v}$ nodes immediately
preceding $v$ such that the branch $vw$ is not on the
resonance paths $\QQ$.
 
Given a set of momenta
and fixed a node $\bar v\in\th^0$, we define the change of variables
$U^{\s_w}_{\bar v w}:\,\ZZZ^\ell\,\otto\,\ZZZ^\ell$,
where $w\in\BB_{\bar v}$,
by fixing a sign $\s_w=\pm1$ and defining
$U^{\s_w}_{\bar v w}(\{\nn_x\})= \{\nn'_x\}$ as:
$$\eqalign{
\nn'_z=&\s_w\nn_z,\qquad z\ge w \; , \cr
\nn'_z=&\nn_z,\qquad {\rm for\ all\ other\ } z\ne \bar v \; , \cr
\nn'_{\bar v}=&\nn_{\bar v} +(1-\s_w)\sum_{z\le w}\nn_z\=\nn_{\bar v}+
(1-\sigma_w)\nn_{\l_w}
\; , \cr} \Eq(4.2)$$
so that, for any choice of the subset $\BB_{1\bar
v}\subseteq\BB_{\bar v}$ of nodes immediately preceding $\bar v$,
there are cancellations which allow us to write
$$ \eqalign{
\sum_{\{\sigma_w\}_{w\in B_{1\bar v}}} &
\RR\hbox{Val}(\th^0,\prod_{w\in\BB_{{1\bar v}}}
U^{\s_w}_{{\bar v}w}\{\nn_x\}) \; \=
\cr &
\= \Big(
\prod_{w\in\BB_{1\bar v}} \int_1^0 dt_{w} \Big) \sum_{||{\V m_{\bar
v}}||=p_{\bar v}} \Big( \prod_{w\in\BB_{1\bar v}} \fra{\dpr}{\dpr
t_{w}} \Big)
\Big( f_{j\nn_{\bar v}(\V t_{\bar v})}\,
(\nn_{\bar v}(\V t_{\bar v}))^{\mm_{\bar v}} \Big) \cdot
\cr &
\cdot \RR \Big\{ {1 \over
\oo_0\cdot\nn_{\l_{\bar v}} } {\rm Val}'(\th^0)  \Big\}
\; , \cr}\Eq(4.3)$$
where:\\
(i) ${\rm Val}'(\th^0)$ is a tensor containing all the other value
factors relative to nodes $v$'s different from $\bar v$.\\
(ii) The free indices of the $p_{\bar v}$-order tensor
$\nn_{\bar v}(\V t_{\bar v})^{\V m_{\bar v}}$
are contracted (by performing the
$\sum_{\V m_{\bar v}}$) with the ones that appear in the tensor
${\rm Val}'(\th^0)$, and ${\V m_{\bar v}}$ is a $\ell$
dimensional positive integer components
vector (with $||{\V m_{\bar v}}||$ denoting the sum of the components) and,
given a vector ${\V b}$, we put ${\V b}^{\V m_{\bar v}}
= b_1^{m_{\bar v 1}}\ldots
b_{\ell}^{m_{\bar v \ell}}$; furthermore
$\V t_{\bar v}=(t_{w_1},\ldots,t_{w_{|\BB_{1\bar v}|}})$ and
$\nn_{{\bar v}}(t_{w_1},\ldots,t_{w_{|\BB_{1\bar v}|}})\=
\nn_{\bar v}(\V t_{\bar v})$ is defined as
$$\nn_{\bar v}(\V t_{\bar v})=
\nn_{\bar v}(t_{w_1},\ldots,t_{w_{|\BB_{{1\bar v}}|}}) = \nn_{\bar v} +
\sum_{w\in \BB_{1\bar v}} 2 t_w\; \nn_{\l_w}=\nn_{\bar v} +
\sum_{w\in \BB_{1\bar v}}
t_w \,\big(\sum_{z\le w}2\nn_z\big) \; , \Eq(4.4) $$
where $\nn_{\bar v}(\V t_{\bar v}) = \nn_{\bar v}$ if $\BB_{1\bar v}=
\emptyset$.\\
(iii) The assumed form \equ(1.5) of the $f_\nn$ allows us to think that
$f_\nn$ is defined on $\RRR^\ell$ rather than on $\ZZZ^\ell$ and hence
to give a meaning to the derivatives of $f_{\nn_v(\V t_v)}$.\\
(iv) We use here and henceforth that $\RR$ acts {\it only}
on the product of propagators $D(\th)$ (see \equ(2.5) and \equ(3.16)),
and the fact that the definition of $\BB_{v}$ after \equ(4.1)
yields that all real resonances remains such under the action
of the change of variables \equ(4.2).
 
\*
 
\0{\bf 4.2.} {\cs Remark.}
The cancellations are expressed by the fact that the change of
variables \equ(4.2) leaves unchanged each factor
$(\nn_{v'}\cdot\ff_{\nn_v})[\oo_0\cdot\nn_{\l_v}]^{-1}$,
except for the nodes $w$ and $\bar v$, whose factors are
modified in the following way:
$$ \eqalign{
{ \nn_{\bar v} \cdot\ff_{\nn_w} \over \oo_0\cdot\nn_{\l_w} }
\quad & \to \quad
- { (\nn_{\bar v}+\zz_w) \cdot\ff_{\nn_w} \over \oo_0\cdot\nn_{\l_w} }
\; , \cr
{ \nn_{\bar v'} \cdot \ff_{\nn_{\bar v}} \over
\oo_0\cdot\nn_{\l_{\bar v}} }
\quad & \to \quad
{ (\nn_{\bar v'}) \cdot\ff_{\nn_{\bar v}+\zz_w}
\over \oo_0\cdot\nn_{\l_{\bar v}} }
\; , \cr} $$
with $\zz_w=2\nn_{\l_w}$: then, if we set $\zz_w=\V0$, the sum
of the two graph values cancel exactly.
 
\*
 
\0{\bf 4.3.} We can study the sum
$S_k(\th^0)=\sum_\nn|\nn|^s|\RR W(\th^0,\nn)|$, where
$\RR W(\th^0,\nn)$ is defined as in \equ(2.6) with
$\hbox{Val}(\th)$ replaced with $\RR \hbox{Val}(\th)$.\annota{1}{
The reader familiar with [BGGM] can skip the following discussion,
which is essentially identical to the one in [BGGM], \S 4,
and leap directly to the final expression \equ(4.27) in \S 4.7.}
We have
$$ \eqalign{
S_k(\th^0) & =
\sum_{\nn} |\nn|^s \Big|\RR W(\th^0,\nn)\Big| \cr & =
\sum_{\nn} |\nn|^s \Big|
f_{j\nn_{v_1}}\,(\nn_{v_1})^{\mm_{v_1}} \,
\RR \Big\{ { 1 \over \oo_0\cdot\nn }
\prod_{w\in\BB_{v_1}}
\Big[ \sum_{\nn_{\l_w}}
W(\th^0_{v_1w},\nn_{\l_w})\Big] \Big\} \Big| \; , \cr} \Eq(4.5)$$
where $v_1$ is the highest node and $\nn_{v_1}=\nn-\sum_{w\in \BB_{v_1}}
\nn_{\l_w}$.
 
Fixed $\nn$ and $\{\nn_{\l_w}\}_{w\in\BB_{v_1}}$
let $h_{v_1}=h_{v_1}(\nn,\{\nn_{\l_w}\}_{w\in \BB_{v_1}})$ be {\sl the
scale of $\nn_{v_1}$}: \ie $\nn_{v_1}$ is such that
$2^{h_{v_1}-1}\leq|\nn_{v_1}|<2^{h_{v_1}}$.  Given $w$ with $w'=v_1$ we
say that $w$ is {\sl out of order} with respect to $v$ if
$$ 2^{h_{v_1}} > 2^op_{v_1}|\nn_{\l_w}| \; ,\qquad o=5 \; , \Eq(4.6)$$
where $p_v$ is the number of branches entering $v$. We denote
$\BB_{1v_1}\=\BB_{1v_1}(\nn,\{\nn_{\l_w}\}_{w\in\BB_{v_1}})
\subseteq\BB_{v_1}$ the nodes $w\in\BB_{v_1}$ which are out of order
with respect to $v_1$.  The number of elements in $\BB_{1v_1}$ will be
denoted $q_v=|\BB_{1v_1}|$.  The notion of $w$ being out of order with
respect to $v_1$ depends on $\{\nn_{\l_w}\}_{w\in B_{v_1}}$ and
$\nn$.
 
Given a set $\{\nn_{\l_w}\}_{w\in B_{v_1}}$ for all choices of
$\sigma_w=\pm 1$ we define the transformation
$$ U(\{\nn_{\l_w}\}_{w\in B_{v_1}})\=
\{\sigma_w\nn_{\l_w}\}_{w\in B_{v_1}} \; , \Eq(4.7) $$
and given a set $C\subseteq \BB_{v_1}$ we call $\UU(C)$ the set of all
transformations $U$ such that $\sigma_w=1$ for $w\not\in C$.
 
If $[2^{h-1},2^h)$ is a scale interval $I_h$, $h=1,2,\ldots$ we call\\
$\bullet$ the first quarter of $I_h$ the {\sl lower part}
$I^-_h=[2^{h-1},\fra542^{h-1})$ of $I_h$,\\
$\bullet$ the fourth quarter of $I_h$
the {\sl upper part} $I^+_h=[\fra782^h,2^h)$ of $I_h$, and\\
$\bullet$ the remaining part the {\sl central part} $I_h^c$.
 
We group the set of branch momenta $\{\nn_{\l_w}\}_{w\in\BB_{v_1}}$ into
collections by proceeding iteratively in the way described below.
The collections will be built so that in each collection the
cancellation discussed in Remark 4.2 above can be exhibited.
 
Fixed $\nn$ and $h$ choose $\{\nn^1_{\l_w}\}_{w\in\BB_{v_1}}$ such that
$|\nn^1_{v_1}|\in I_{h}^c$: such $\{\nn^1_{\l_w}\}_{w\in\BB_{v_1}}$
is called a {\sl representative}. Given the representative we define\\
$\bullet$
the {\sl branch momenta collection} to be set of
the $\{\nn_{\l_w}\}_{w\in \BB_{v_1}}$ of the form
$$U(\{\nn^1_{\l_w}\}_{w\in\BB_{v_1}}),\qquad U\in\UU(\BB_{1v_1}(\nn,
\{\nn^1_{\l_w}\}_{w\in\BB_{v_1}})) \; ; \Eq(4.8)$$
\\$\bullet$
the {\sl external momenta collection} to be the set of momenta
$$\nn^{1U}_{v_1}=\nn-\sum_{w\in \BB_{v_1}}\s_w\nn^1_{\l_w},\qquad{\rm for}
\ U\in\UU(\BB_{1v_1}(\nn, \{\nn^1_{\l_w}\}_{w\in\BB_{v_1}}))\;.\Eq(4.9) $$
Note that the elements of the above constructed external momenta
collection need not be necessarily contained in $I_{h}^c$.
 
\*
 
We consider then another representative
$\{\nn^2_{\l_w}\}_{w\in\BB_{v_1}}$ such that $|\nn^2_{v_1}|\in I_h^c$
{\it and} not belonging to the branch momenta collection associated with
$\{\nn^1_{\l_w}\}_{w\in\BB_{v_1}}$, if there are any left; and we
consider the corresponding branch momenta and external momenta
collections as above. We proceed in this way until all the
representatives such that $\nn_1$ is in $I_{h}^c$, for the given $h$,
have been put into some collection of branch momenta.
 
We then repeat the above construction with the interval $I^-_{h}$
replacing the $I^c_h$, always being careful not to consider
representatives $\{\nn_{\l_w}\}_{w\in\BB_{v_1}}$ that appeared as
members of previously constructed collections.  It is worth pointing out
that not all the external momenta $\nn^U_{v_1}$, $U\in
\UU(\BB_{1v_1}(\nn,\{\nn^1_{\l_w}\}_{w\in\BB_{v_1}}))$, are in $I_h^-$,
but they are all in the corridor $I_{h-1}^+\cup I_h^-$, by \equ(4.6).
 
Finally we consider the interval $I^+_{h-1}$, (if $h=1$ we simply skip
this step). The construction is repeated for such intervals.
 
Proceeding iteratively in this way starting from $h=1$ and, after
exhausting all the $h=1$ cases, continuing with the $h=2,3\ldots$ cases,
we shall have grouped the sets of branch momenta into collections
obtainable from a representative $\{\nn_{\l_w}\}_{w\in\BB_{v_1}}$ by
applying the operations $U\in \UU(\BB_{1v_1}(\nn,
\{\nn^1_{\l_w}\}_{w\in\BB_{v_1}}))$ to it.
{\it Note that, in this way, when the interval $I^+_{h-1}$ is considered,
all the remaining representatives are such that
$|\nn^U_{v_1}|\in I_{h-1}^+$ for all $U\in
\UU(\BB_{1v_1}(\nn,\{\nn^1_{\l_w}\}_{w\in\BB_{v_1}}))$.}
 
Note that the graphs with momenta in each collection
are just the graphs involved in the parity cancellation described in the
previous section. In fact if $U$ is generated by the signs
$\{\s_w\}_{w\in\BB_v}$, we have
$$ \eqalign{
& \nn_{v_1}^U=\Big(\prod_{w\in\BB_{{1v_1}}}
U^{\s_w}_{{v_1}w}\{\nn_x\}\Big)_{v_1} , \cr
& (U(\{\nn_{\l_{\tilde
w}}\}_{\tilde w\in B_{v_1}}))_w=\sum_{z\leq w}\Big (\prod_{\tilde
w\in\BB_{{1v_1}}} U^{\s_{\tilde w}}_{{v_1}\tilde w}
\{\nn_x\}\Big)_{z} , \cr} \Eq(4.10) $$
where, given the sets $\{\nn_x\}$ and $\{\nn_{\l_{\tilde w}}\}$,
$(\{\nn_x\})_v$ denotes the external momentum in $\{\nn_x\}$
corresponding to the node $v$ and
$(\{\nn_{\l_{\tilde w}}\})_w$ denotes the branch momentum in
$\{\nn_{\l_{\tilde w}}\}$ corresponding to the branch $\l_w$.
 
\*
 
\0{\bf 4.4.} {\cs Remark.}
The complexity of the above construction is due to the
necessity of avoiding overcountings. In fact it is possible that, for some
$U\in \UU(\BB_{1v_1}(\nn,\{\nn_{\l_w}\}_{w\in\BB_{v_1}})$, one has
$$\BB_{1v_1}(\nn,U(\{\nn_{\l_w}\}_{w\in\BB_{v_1}}))\not=
\BB_{1v_1}(\nn,\{\nn_{\l_w}\}_{w\in\BB_{v_1}}) \; , \Eq(4.11) $$
because the scale of $\nn^U_{v_1}$ may be $h-1$, while that of $\nn_{v_1}$
may be $h$; so that if one considered, for instance, $I^+_{h-1}$ before
$I^-_{h}$ overcountings would be possible, and in fact they would
occur.
 
\*
 
\0{\bf 4.5.} A convenient way to rewrite \equ(4.5) is the following:
$$\eqalign{
\sum_\nn & |\nn|^s \Big|\sum_{h_{v_1}}
\sum^*_{\{\nn_{\l_w}\}_{w\in\BB_{v_1}}}\sum_{U\in
\UU(B_{1v_1})} f_{j\nn_{v_1}^U} \, (\nn_{v_1}^U)^{\mm_{v_1}} \, \cdot \cr
\cr & \cdot \, \RR \Big\{
{1\over \oo_0\cdot\nn } \prod_{w\in\BB_{v_1}}
W(\th^0_{v_1w},\s_w\nn_{\l_w}) \Big\} \Big| \; , \cr} \Eq(4.12)$$
where $\sum^*_{\{\nn_{\l_w}\}_{w\in\BB_{v_1}}}$ means sum over
the above defined representatives such that $\nn_{v_1}$ is compatible
with $h_{v_1}$; and we abridge
$\BB_{1v_1}(\nn,\{\nn_{\l_w}\}_{w\in\BB_{v_1}})$ by $\BB_{1v_1}$ in
conformity with the notations introduced after \equ(4.6).
The explicit sum over the scales $h_{v_1}$ is introduced to simplify the
bounds analysis that we perform later, see \S4.8.  Note that
$\nn_{v_1}^U$ is, in general, not compatible with $h_{v_1}$, \ie we are
grouping together also terms with different scale label (but the
difference in scale is at most one, see \equ(4.16) below).
 
Noting that, by the parity properties of $\ff$,
$$ W(\th^0_{v_1w}, \s_w\nn_{\l_w})=\sigma_wW(\th^0_{v_1w},\nn_{\l_w})
\Eq(4.13) $$
we have, from \equ(4.12),
$$ \eqalign{
\sum_\nn & |\nn|^s \Bigl|\sum_{h_{v_1}}
\sum^*_{\{\nn_{\l_w}\}_{w\in\BB_{v_1}}} \sum_{U\in
\UU(\BB_{1v_1})} f_{j\nn_{v_1}^U} \, (\nn_{v_1}^U)^{\mm_{v_1}} \, \cdot
\cr & \cdot \,
\RR \Big\{ \Big( \prod_{w\in\BB_{1v_1}}\sigma_w\Big)
{ 1\over \oo_0\cdot\nn }
\prod_{w\in\BB_{v_1}}W(\th^0_{v_1w},\nn_{\l_w})
\Big\} \Bigr| \; . \cr} \Eq(4.14) $$

We can apply the interpolation in \equ(4.3) to the node $v$ and rewrite
\equ(4.14) as
$$ \eqalign{
\sum_\nn & |\nn|^s  \Big|\sum_{h_{v_1}}
\sum^*_{\{\nn_{\l_w}\}_{w\in\BB_{v_1}}}
\Big[ \sum_{||\mm_{v_1}||=p_{v_1}} \Big( \prod_{w\in
\BB_{1v_1}}\ig_1^0 dt_w \Big) \cdot
\cr &\cdot
\Big( \prod_{w\in \BB_{1v_1}}\fra{\dpr}{\dpr t_w} \Big) \Big(
f_{j\nn_{v_1}(\V t_{v_1})} \,
\big( \nn_{v_1}(\V t_{v_1})\big)^{\mm_{v_1}}
\Big) \Big] \cdot \cr & \cdot \, \RR \Big\{  {1 \over \oo_0\cdot\nn }
\prod_{w\in\BB_{v_1}}W(\th^0_{v_1w},\nn_{\l_w})
\Big\} \Bigl| \; , \cr} \Eq(4.15) $$
where if $\BB_{1v_1}=\emptyset$ no interpolation is made; and we note
that by \equ(4.3), by the definition of nodes out of order
and by the iterative grouping of the representatives:
$$
2^{h_{v_1}-2}\le|\nn_{v_1}(\V t_{v_1})| < 2^{h_{v_1}}\; ,\Eq(4.16) $$
so that the interpolation formulae discussed in \S 4.1 {\it can be
used} because no singularity arises in performing the $\V
t_{v_1}$-integrations.
 
By the definition of $W(\th^0, \nn)$ we can write \equ(4.15) as
$$\eqalign{
\sum_\nn & |\nn|^s \Big|\sum_{h_{v_1}}
\sum^*_{\{\nn_{\l_w}\}_{w\in\BB_{v_1}}}
\Big[ \sum_{||\mm_{v_1}||=p_{v_1}} \Big(
\prod_{w\in\BB_{1v_1}}\ig_1^0 dt_w \Big)
\cdot\cr
&\Big(
\prod_{w\in\BB_{1v_1}}
\fra{\dpr}{\dpr t_w} \Big) \Big(
f_{j\nn_{v_1}(\V t_{v_1})} \,
\big(\nn_{v_1}(\V t_{v_1})\big)^{\mm_{v_1}}
\Big)  \Big]
\cr
& \cdot \RR \Big\{ { 1 \over \oo_0\cdot\nn }
\prod_{w\in\BB_{v_1}}
\sum_{[\{\nn_x\}_{x\leq w};\,\nn_{\l_w}]}
\Val(\th^0_{v_1w},\{\nn_x\}_{x \leq w}) \Big\}
\Bigl| \; , \cr} \Eq(4.17) $$
where the sum over $[\nnx_{x\le w};\,\nn_{\l_w}]$ is a sum over the
$\nnx_{x\le w}$ with $\sum_{x\le w}\nn_x=\nn_{\l_w}$.
 
If we use (see \equ(4.3))
$$\fra{\dpr}{\dpr t_w}\=\Big( 2\nn_{\l_w}
\cdot\fra{\dpr}{\dpr\nn}\Big)_{\nn=\nn_v(\V t_v)} \=\Big( \sum_{z\le w}
2\nn_z\cdot\fra{\dpr}{\dpr\nn}\Big)_{\nn=\nn_v(\V t_v)} \; , \Eq(4.18)$$
to compute differentiations with respect to $t_w$, we can write
\equ(4.15) as
$$\eqalign{
\sum_\nn & |\nn|^s \Big|\sum_{h_{v_1}}
\sum^*_{\{\nn_{\l_w}\}_{w\in\BB_{v_1}}}
\Big\{\sum_{||\mm_{v_1}||=p_{v_1}} \Big( \prod_{w\in\BB_{1v_1}}
\int_1^0 dt_w \Big) \cdot
\cr
&
\cdot \Big( |\bar \nn|^{-(p_{v_1}-q_{v_1})}
\fra{\dpr^{q_{v_1}}}{\dpr \bar\nn^{q_{v_1}}}
\, f_{j\bar\nn} \, \big( \bar \nn \big)^{ \mm_{v_1}}
\Big)_{\bar \nn =\nn_{v_1}(\V t_{v_1})} \Big\} \cdot
\cr
&
\cdot \Big[ \prod_{w\in\BB_{1v_1}} \Big( \sum_{z \le w} 2\nn_z \Big)
\Big] \Big[ \prod_{w\in\BB_{v_1}\setminus\BB_{1v_1}} | \nn_{v_1}(\V
t_{v_1}) | \Big]
\cr
& \RR \Big\{ {1\over \oo_0\cdot\nn }
\prod_{w\in \BB_{v_1}}\Big[\sum_{[\{\nn_x\}_{x\leq
w};\,\nn_{\l_w}]} \Val(\th_{v_1w},\{\nn_x\}_{x \leq w})\Big]
\Big\} \Bigl|
\; , \cr} \Eq(4.19)$$
where we recall that $q_{v_1}=|\BB_{1v_1}|$ and the sum over $[\nnx_{x\le
w};\,\nn_{\l_w}]$ again denotes sum over the $\nnx_{x\le w}$ with
$\sum_{x\le w}\nn_x=\nn_{\l_w}$; here the factor
$|\bar\nn|^{-(p_{v_1}-q_{v_1})}$ (which, computed for
$\bar\nn=\nn_{v_1}(\V t_{v_1})$, is identical to the inverse of
$[\prod_{w\in\BB_{v_1}\setminus\BB_{1v_1}}$ $|\nn_{v_1}
(\V t_{v_1})|]$) has been
introduced so that a dimensional estimate of the factor in the second
line of \equ(4.19) can be taken proportional to $a\,2^{-h_{v_1}b}$
(see the end of \S 1.3 and \equ(4.16)).
 
If $w\in\BB_{v_1}\setminus\BB_{1v_1}$ we have
$$ \eqalign{
\prod_{w\in\BB_{v_1}\setminus\BB_{1v_1}} | \nn_{v_1}(\V t_{v_1}) |
& = \prod_{w\in\BB_{v_1}\setminus\BB_{1v_1}}
(2^o p_{v_1})\,{\bf\tilde x}_{v_1w}(\V t_{v_1})
\cdot\nn_{\l_w} \cr
& =\prod_{w\in\BB_{v_1}\setminus\BB_{1v_1}} (2^o p_{v_1}) \,
{\bf\tilde x}_{v_1w} (\V t_{v_1})\cdot
\sum_{z\le w} \nn_z \; , \cr} \Eq(4.20) $$
where $\tilde \V x_{v_1w}(\V t_{v_1})$ is a suitable vector depending on
$\nn_{\l_w}$ {\it but not on the individual terms $\nn_z$}, and such
that $|{\bf \tilde x}_{vw}(\V t_{v_1})|<1$.
 
We obtain, with the above notations (and taking $o=5$, see \equ(4.6)),
$$\eqalign{
S_k(\th^0) & = \sum_\nn |\nn|^s \Big|\sum_{h_{v_1}}
\sum^*_{\{\nn_{\l_w}\}_{w\in\BB_{v_1}}}
\Big\{\sum_{||\mm_{v_1}||=p_{v_1}}
\Big( \prod_{w\in\BB_{1v_1}}\int_1^0 dt_w \Big)\cdot
\cr
&
\cdot \Big( \fra{Y_{v_1}(\V t_{v_1})}{|\bar\nn|^{p_{v_1}-q_{v_1}}}
\fra{\dpr^{|\BB_{1v_1}|}}{\dpr \bar\nn^{|\BB_{1v_1}|}}
\, f_{j\bar\nn} \, \big( \bar\nn \big)^{\mm_{v_1}}
\Big)_{\bar\nn=\nn_{v_1}(\V t_{v_1})}\Big\}
\Big[\prod_{w\in \BB_{v_1}} \Big( \sum_{z \le w} 2\nn_z \Big) \Big] \cdot
\cr
& \RR \Big\{ {1\over\oo_0\cdot\nn}
\prod_{w\in \BB_{v_1}} \Big[ \sum_{\{\nn_x\}_{x\leq w};\,\nn_{\l_w}}
\Val(\th_w,\{\nn_x\}_{x \leq w}) \Big] \Big\} \Big| \; , \cr}\Eq(4.21) $$
where the tensor
$$Y_{v_1}(\V t_{v_1})=\prod_{w\in\BB_{v_1}\setminus\BB_{1v_1}}
2^4p_{v_1}{\bf\tilde x}_{v_1w}(\V t_{v_1}) \Eq(4.22)$$
depends also on $\nn$ and $\{\nn_{\l_w}\}_{w\in\BB_{v_1}}$ (although
this dependence is not shown, to simplify the notation), and
has to be contracted with the external momenta
$\nn_z$, $z\le w\in\BB_{1v_1}$.
 
\*
 
\0{\bf 4.6.}
Developing the sum $ \sum_{z \le w} 2\nn_z $ in \equ(4.21) $S_k(\th^0)$ is
given by a sum of terms corresponding to a collection of nodes lying on
the paths $P({v_1},z({v_1},w))$ leading from ${v_1}$ to a node $z$: the
collection is defined by the ``choices" of one particular addend
$2\nn_z$ in the sum $ \sum_{z \le w} 2\nn_z $, with $z=z(v_1,w)$,
$w\in\BB_{v_1}$. Therefore, in general, we can think that \equ(4.21)
corresponds to a sum over a collection of paths $P({v_1},z({v_1},w))$ for
the $w\in\BB_{{v_1}}$. The paths are regarded as totally ordered (and
gapless) sequences of nodes on $\th_0$.
 
We can call $\PP_1$ the family of the possible collections of paths that
arise when expanding the sums $\sum_{z\le w}$ in \equ(4.21): each
element $\V P_1$ of $\PP_1$ can be identified with one contribution to
\equ(4.21).  And, by using the notation $\V
t_{v}=\{t_w\}_{w\in\BB_{1v}}$ as in \equ(4.3),
the result is the following more
explicit interpolation formula reexpressing the r.h.s. of \equ(4.21)
$$\eqalign{
\sum_\nn & |\nn|^s \Big|\sum_{h_{v_1}}\sum_{\V
P_1 \in \PP_1}
\sum^*_{\{\nn_{\l_w}\}_{w\in\BB_{v_1}}} \Big\{\sum_{||
\mm_{v_1}||=p_{v_1}} \Big(
\prod_{w\in\BB_{1v_1}} \int_1^0
dt_w \Big) \cdot
\cr
&
\cdot \Big(\fra{Y_{v_1}(\V t_{v_1})}{|\bar\nn|^{p_{v_1}-q_{v_1}}}
\fra{\dpr^{q_{v_1}}}{\dpr \bar \nn^{q_{v_1}}}
f_{j\bar\nn} \, \big( \bar\nn
\big)^{\mm_{v_1}} \Big)_{\bar\nn=\nn_{v_1}(\V t_{v_1})} \Big\}
\cdot \Big( \prod_{z:P({v_1},z)\in \V P_1}
2\nn_z \Big) \cdot
\cr &
\cdot \RR \Big\{ {1\over\oo_0\cdot\nn}
\prod_{w\in \BB_{v_1}} \Big[ \sum_{\{\nn_x\}_{x\leq w};\,\nn_{\l_w}}
 \Val(\th^0_{v_1w},\{\nn_x\}_{x \leq w})\Big) \Big] \Big\} \Big|
\; , \cr} \Eq(4.23)$$
where the interpolation is considered when $\BB_{1v_1}\ne\emptyset$ (\ie
when it makes sense), and the indices have to be contracted suitably.
 
The above formula can be rewritten as
$$\eqalign{
\sum_{\nn} & |\nn|^s \Big|\sum_{h_{v_1}}
\sum^*_{\{\nn_{\l_w}\}_{w\in \BB_{v_1}}}
\sum_{\V P_1\in \PP_1} \Big(\prod_{v\in[\V P_1]}
\sum_{\{\nn_{\l_y}\}_{y\in \BB_{v}}}\Big)
\cr
& \RR \Big\{\sum_{||\mm_{v_1}||=p_{v_1}}
\Big( \prod_{w\in\BB_{1v_1}}\ig_1^0 dt_w \Big) \cdot
\Big( \fra{Y_{v_1}(\V t_{v_1})}{|\bar\nn|^{p_{v_1}-q_{v_1}}}
\fra{\dpr^{q_{v_1}}}{\dpr \bar\nn^{q_{v_1}}}
{ f_{j\bar\nn} \, \big( \bar\nn \big)^{\mm_{v_1}} \over
\oo_0\cdot\nn}
\Big)_{\bar \nn=\nn_{v_1}(\V t_{v_1})} \,
\cr
&
\cdot\prod_{v\in [\V P_1]} \sum_{||\mm_{v}||=p_{v}}
(2\nn_v)^{\h_v}
\, { \ff_{\nn_v}^{\V 1} \, (\nn_v)^{\mm_v+1} \over
\oo_0\cdot\nn_{\l_v} }
\prod_{y\in \BB_v/[\V P_1]}
W(\th^0_{vy},\nn_{\l_y}) \Big\} \Big| \; , \cr}\Eq(4.24)
$$
where\\
$\bullet$ $[\V P_1]=\bigcup_{w\in\BB_{v_1}}P(v_1,z(v_1,w))/\{v_1\}$,\\
$\bullet$ $f_{\nn_v}^{\V 1} = f_{1\nn_v}^{1_1}\ldots
f_{\ell\nn_{v}}^{1_{\ell}}$, with $\|{\V 1}\|=1$, is
contracted with a factor in $(\nn_{v'})^{\mm_{v'}}$, and\\
$\bullet$ $\h_v$ is equal to $1$ if $v=z(v_1,w)$
for some $w\in\BB_{v_1}$ and $0$ otherwise.
 
We are now in position to iterate the resummation done in the previous
section leading from \equ(4.5) to \equ(4.21) and ``concerning" the
highest node $v_1$.  For each $\tilde v\in\V P_1$, $\tilde v < v_1$,
let $h_{\tilde
v}=h_{\tilde v}(\nn_{\l_{\tilde v}}, \{\nn_{\l_w}\}_{w\in \BB_{\tilde
v}})$ be {\sl the scale of $\nn_{\tilde v}$}, \ie $\nn_{\tilde
v}=\nn_{\l_{\tilde v}}-\sum_{w\in \BB_{\tilde v}}\nn_{\l_w}$ is such
that $2^{h_{\tilde v}-1}\leq|\nn_{\tilde v}|<2^{h_{\tilde v}}$.
 
Given an immediate predecessor $w$ of $\tilde v$ we say that $w$ is
{\sl out of order} with respect to $\tilde v$ if
$$ 2^{h_{\tilde v}} > 2^5p_{\tilde v}|\nn_{\l_w}| \;  , \Eq(4.25)$$
where $p_{\tilde v}$ is the number of branches entering $\tilde v$.  We
denote $\BB_{1\tilde v}\=\BB_{1\tilde v}(\nn_{\l_{\tilde v}},
\{\nn_{\l_w}\}_{w\in\BB_{\tilde v}}) \subseteq\BB_{\tilde v}$ the nodes
$w\in\BB_{\tilde v}$ which are out of order with respect to $\tilde v$.
 
Given a set $\{\nn_{\l_w}\}_{w\in B_{\tilde v}}$ for all choices of
$\sigma_w=\pm 1$ we define
$$U(\{\nn_{\l_w}\}_{w\in B_{\tilde v}})\=
\{\sigma_w\nn_{\l_w}\}_{w\in B_{\tilde v}} \; , \Eq(4.26)$$
and given a set $C\subseteq \BB_{\tilde v}$
we call $\UU(C)$ the set of all transformations such that $\sigma_w=1$ for
$w\not\in C$.
 
We group the set of branch momenta $\{\nn_{\l_w}\}_{w\in\BB_{\tilde v}}$
and the external momenta into collections by proceeding, very closely
following the preceding construction, with $\nn_{\l_w}$ playing the role of
$\nn$, in the way described below.
 
Fixed $\nn_{\l_{\tilde v}}$ and $h$ we choose a
$\{\nn^1_{\l_w}\}_{w\in\BB_{\tilde v}}$ such that $|\nn^1_{\tilde v}|\in
I_{h}^c$ where $\nn^1_{\tilde v}=\nn_{\l_{\tilde
v}}-\sum_{w\in\BB_{\tilde v}}\nn_{\l_w}$.
 
Then $\{\nn^1_{\l_w}\}_{w\in\BB_{\tilde v}}$ is called
a {\sl representative}.  For such representative we define the
{\sl branch momenta collection},
associated with it to be the set of the
$\{\nn_{\l_w}\}_{w\in \BB_{\tilde v}}$ having the form
$U(\{\nn^1_{\l_w}\}_{w\in\BB_{\tilde v}})$ and the
{\sl external momenta collection}
to be the set of momenta $\nn^{1U}_{\tilde v}=\nn-\sum_{w\in
\BB_{\tilde v}}\s_w\nn^1_{\l_w}$, for $U\in\UU(\BB_{1\tilde
v}(\nn_{\l_{\tilde v}}, \{\nn^1_{\l_w}\}_{w\in\BB_{\tilde v}})/[\V
P_1])$.  Note again that the above constructed external momenta
collection is not necessarily contained in $I_{h}^c$.
 
We consider then another representative
$\{\nn^2_{\l_w}\}_{w\in\BB_{\tilde v}}$ such that $|\nn^2_{\tilde v}|\in
I^c_h$ and does not belong to the just constructed branch momenta
collection associated with $\{\nn^1_{\l_w}\}_{w\in\BB_{\tilde v}}$, if
there is any; and then we consider the branch momenta collections and
external momenta collections obtained from
$\{\nn^2_{\l_w}\}_{w\in\BB_{\tilde v}}$ by the corresponding  $U$
transformations.  And, as previously done, we proceed in this way until
all the representatives such that $\nn^1_{\tilde v}$ is in $I_{h}^c$ are
in some external momenta collections.
 
The construction is repeated for the interval $I^-_{h}$, always
being careful not to consider $\{\nn_{\l_w}\}_{w\in\BB_{\tilde v}}$ that
have been already considered, and finally for the interval $I^+_{h-1}$,
see \S 4.3.
 
Proceeding iteratively in this way and considering the same sequence of
$h$'s as in the previous case (\ie the natural $h=1,2,\ldots$), at the
end we shall have grouped the set of branch momenta into collections
obtainable from a representative $\{\nn_{\l_w}\}_{w\in\BB_{\tilde v}}$
by applying the operations $U\in \UU(\BB_{1\tilde v}(\nn_{\l_w},
\{\nn^1_{\l_w}\}_{w\in\BB_{\tilde v}})\setminus [{\bf P}_1])$ to it.
 
{\it In other words the definition of the representatives
$\{\nn_{\l_w}\}_{w\in\BB_{\tilde v}}$ is identical to the one for $v_1$
except that the collections are defined {\it only} by transformations
changing the branch momentum of the lines emerging from the nodes
in $\BB_{1\tilde v}$ {\it but not in $\V P_1$}.}
 
We repeat the above construction for all $\tilde v\in \V P_1$ until all
the $\tilde v\in \V P_1$ are considered starting from the $\tilde v$ with
$\tilde v'=v$ and, after exhausting them, continuing with $\hat v$ with
$\hat v'=\tilde v$ and so on. We call $\BB_{\tilde v}(\V P_1)$
the nodes $w$ immediately preceding $\tilde v$ {\it but which are not
on the union of the paths $P\in \V P_1$}, and $\BB_{1\tilde v}(\V P_1)$
the nodes in $\BB_{\tilde v}(\V P_1)$ which are out of order with
respect to $\tilde v$; the set of just described
transformations will be denoted by ${\cal U}(\BB_{1\tilde v}(\V P_1))$.
 
Proceeding as we did for the highest node $v_1$ and by performing the
analogues of the transformations leading from \equ(4.15) to \equ(4.24),
we construct for each $\tilde v\in \V P_1$ new paths $\V P_2$ which, by
construction, {\it will not have common branches} with those in $\V
P_1$; call $\PP_2$ the collection of the pairs $\V P_1,\V P_2$.  The
crucial point is that the factors ${\bf\tilde x}_{vw}(\V t_v)$ are the
same for all the terms generated by the action of $U\in {\cal
U}(\BB_{1v}(\V P_1))$, by \equ(4.20). We iterate then this procedure.
 
Eventually we end up by constructing a {\sl pavement} $\V P$ of the
graph with non overlapping paths (and the union of the paths does cover
the graph); note that the paths are ``ordered'', in the
sense that they are formed only by comparable lines.
 
We call $\PP$ the collection of all such pavements; $\BB_v(\V P)$, $\V
P\in\PP$, will be the set of nodes $w$ immediately preceding $v$ and
such that a path $P(v,z(v,w))\in\V P$ starting from $v$ passes through $w$,
and $\BB_{1v}(\V P)$ is the collection of nodes in $\BB_v(\V P)$
out of order with respect to $v$. Note that in general
$\BB_v(\V P)\subseteq \BB_v$ (unless $v$ is the highest node $v_1$, when
$\BB_{v_1}(\V P)=\BB_{v_1}$).
 
The set of ``path head" nodes $v$, \ie the upper end nodes of paths in
$\V P$, will be denoted $M_h(\V P)$: hence if $v\not\in M_h(\V P)$ (\ie if no
path in $\V P$ has $v$ as path head) then $\BB_v(\V P)=\emptyset$;
likewise $M_e(\V P)$ will denote the set of ``path end'' nodes,
\ie the nodes $z$ such that $P(v,z)$ is a path in $\V P$.
 
\*
 
\0{\bf 4.7.}
Then we see that \equ(4.22) leads the following {\sl path expansion}
for $S_k(\th^0)$ summarizing our analysis
$$ \eqalignno{
\sum_{\nn}|\nn|^s| & W(\th^0,\nn)|=\sum_\nn |\nn|^s \Big|
\sum_{\{h_x\} }
\sum_{\V P\in\PP} {\sum_{\{\nn_{\l}\}}}^*
\prod_{v \in M_h(\V P)} & \eq(4.27) \cr
&
\Big\{ \Big( \prod_{w\in\BB_{1v}(\V P)} \ig_1^0\,dt_w \Big)
\sum_{||\V m_v||=p_v}\, \OO_v\Big(
\ff_{\nn_v(\V t_v)}^{\V 1} \, (\nn_v(\V t_v))^{\V m_v} \,
Y_v(\V t_v) \Big) \Big\}
\Big| \, \RR D(\th^0) \; , \cr} $$
where\\
(1) $\V P$ is a partial pavement of the graph with non overlapping
{\sl paths} such that: (1.1) a path $P(v,z)$ is a connected set of
comparable lines connecting the node $v$ to the node $z<v$;
(1.2) the resonance paths are not contained in any path;
(1.3) for any line $\l$ which is not contained in any resonance path
there is one path in $\V P$
covering it: $\l\in P(v,z)$ for some $P(v,z)\in \V P$;\\
(2) $M_h(\V P)$ is the collection of upper end nodes of the
paths in $\V P$, and $M_e(\V P)$ of lower end nodes;\\
(3) $\BB_v(\V P)$ is the set of nodes immediately
preceding $v$, and $\BB_{1v}(\V P)$ is the set of nodes
$w\in \BB_v(\V P)$ which are {\sl out of order with respect to
$v$}, \ie such that
$$ 2^{h_v} > 2^5 p_v|\nn_{\l_w}| \; ; \Eq(4.28) $$
\\(4) $Y_v(\V t_v)$ is defined as
$$Y_v(\V t_v)=\cases{\prod_{w\in\BB_v(\V P)\setminus \BB_{1v}(\V
P)}( 2^4p_v\tilde \V x_{vw}(\V t_v)) \; , & if $v\in M_h(\V P) \; , $\cr
1 \; , & otherwise $\; ,$ \cr} \Eq(4.29)$$
if $\tilde \V x_{vw}(\V t_v)$ is the vector
defined via the implicit relation
$$ \nn_v(\V t_v) = 2^5p_v\,\tilde \V x_{vw}(\V t_v) \cdot
\nn_{\l_w} \; , \Eq(4.30) $$
so that $|\tilde \V x_{vw}(\V t_v))|\le 1$ and
$\tilde \V x_{vw}(\V t_v))$ depends on $\nn_{\l_w}$
but not on the individual external momenta which add to
$\nn_{\l_w}$;\\
(5) the operator $\OO_v$ is defined as
$$ \eqalign{
\OO_v & \Big( (\nn_v(\V t_v))^{\V m_v}\,Y_v(\V t_v)\,
\ff_{\nn_v(\V t_v)} \Big) =\cr
&= \Big(\fra{Y_v(\V t_v)}
{|\bar \nn|^{|\BB_v(\V P)|-|\BB_{1v}(\V P)|}}
\fra{\dpr^{|\BB_{1v}(\V P)|}}{{\dpr \bar\nn}^{|\BB_{1v}(\V P)|}}
\ff_{\bar\nn}^{\V 1} \, (\bar\nn)^{\V m_v}\,(2\bar\nn)^{\h_v}
\Big)_{\bar\nn=\nn_v(\V t_v)}
\; , \cr} \Eq(4.31) $$
with $\h_v=1$ if $v\in M_e(\V P)$, and $\h_v=0$ otherwise,
and $\ff_{\nn}^{\V 1}$ defined after \equ(4.24);\\
(6) the sum over $\{\nn_{\l}\}$ has
the restriction that the
external momentum configuration $\{\nn_x\}$ is compatible
with the scales $\{h_x\}$;\\
(7) $\RR D(\th^0)$ is the same for all graphs involved in
the cancellations mechanisms, as the moduli of the momenta
do not change under the action of the change of
variables \equ(4.2), and the signs are taken into account
by the interpolation formula \equ(4.3) (see Remark 4.2).
 
\*
 
\0{\bf 4.8.} We can bound
$$ \eqalign{
|\nn|^s & \prod_{v\in\th^0}
\| \OO_v \Big( \ff_{\nn_v(\V t_v)}^{\V 1}
(\nn_v(\V t_v))^{\V m_v} \, Y_v(\V t_v) \, \Big) \| \le \cr
& \le \prod_{v\in\th^0} D_1 D_2^{p_v} q_v!\,p_v^{p_v-q_v}
\,2^{h_v(1-b+s+\h_v)} \le \prod_{v\in\th^0}
D_3 D_4^{p_v} p_v! \,2^{h_v(1-b+s+\h_v)} \; , \cr} \Eq(4.32) $$
for suitable constants $D_j$, and use
$ \prod_{v\in\th^0}|\nn_v|^{{\h\over2}\t}
\le \prod_{v\in\th^0} 2^{h_v{\h\over2}\t} $,
so that
$$ \eqalign{
&\sum_{\{ \nn\}}|\nn|^s\Big|\sum_{\th^0} \RR W(\th^0,\nn) \Big|\le\cr
& \le \max_{\V P \in \PP} \Big\{{C_3^k}
\Big[ \prod_{v\le v_0} {p_v!}\Big] \sum_{\{h_x\}}
\prod_{v\le v_0} \Big[ 2^{h_{v}(p_v+s+\ell+{\h\over2}\t-b)} \prod_{
P(v,z) \in \V P} 2^{(h_{z}-h_{v})} \Big] \Big\}  \; . \cr}\Eq(4.33)$$

Then, setting $b=2+s+\ell+{\h\over2}\t+\m$, with $\m>0$,
and exploiting the identity
$$ \sum_{v<v_1} h_v p_v = \sum_{v<v_1} h_{v'} \;
, \qquad v_1'=r \; , \Eq(4.34) $$
one obtains for \equ(4.5) the bound
$$\eqalignno{
& \sum_{\{ \nn\}}|\nn|^s\Big|\sum_{\th^0} \RR W(\th^0,\nn) \Big|\le\cr
& \le {C^k\prod_v{p_v!}} \sum_{ \{h_x\}} \Big[
2^{-(2+\m)h_{v_0}} \prod_{v <  v_0} 2^{-(1+\m) h_v}
\prod_{ vw \in \QQ} 2^{h_v-h_{w}} \Big] \; ,& \eq(4.35) \cr }$$
for a suitable constant $C$.
 
We see that there is at most one factor $2^{h_v-h_w}$ per
node $v$, because the resonance paths are totally ordered, so that
the factors $2^{-h_v}$ in $2^{-(1+\m)h_v}$ compensate
(when necessary) the factors $2^{h_v}$ in
$2^{h_v-h_{v'}}$ (and $2^{-h_{v'}}\le 1$): then
the sum over the scales can be performed.
 
There are $k!/\prod_v p_v!$ graphs with given $p_v$'s and fixed
shape ({\sl Cayley's formula}, see [HP]),
so that the sum over the graph orders weighed by $\e^k$ can be
performed if $\e$ is small enough; in particular we obtain that $\hh\in
C^{(s)}(\TTT^{\ell})$, if $f\in \hat C^{(2+s+{\h\over 2}
\t+\m)}(\TTT^{\ell})$, with $\m>0$.
 
\*
 
\0{\bf 4.9.} Then we can pass to the equation \equ(2.10) for $\HH$,
with $\Val^*(\th)$ defined in \equ(2.9).
In such a case we give the extra prescription not to
apply the ultraviolet interpolation procedure to the
path $\CC(v_0,\tilde v)$; equivalently, modify
slightly the definition of the set $\BB_v$ after \equ(4.1):
$\BB_v$ is the the subset of the nodes $w$
among the $p_{v}$ nodes immediately
preceding $v$ such that the branch $vw$ is neither on the
resonance paths $\QQ$ nor on $\CC(v_0,\tilde v)$.
 
Then we obtain again a formula like \equ(4.17),
with respect to which there are the following differences.\\
(1) $\V P$ is the partial pavement such that, besides the conditions
(1.1)$\div$(1.3) after \equ(4.27), verifies the further condition:
(1.4) there is no overlapping between $\CC(v_0,\tilde v)$ and
any $P(v,z)\in\V P$.\\
(2) If $v\in\CC(v_0,\tilde v)$, $\ff_{\nn_v}^{\V 1}$ has to
be contracted with a factor in $(\nn_{v'{'}})^{\mm_{v'{'}}}$,
where $v'{'}\in\CC(v_0,\tilde v)$ is the node
on $\CC(v_0,\tilde v)$ immediately preceding $v$.\\
(3) If $v\in\CC(v_0,\tilde v)$, the factors $(\nn_v)^{\mm_v}$
arise from the $p_v-1$ branches not contained in
$\CC(v_0,\tilde v)$ and entering $v$ and from the branch
on $\CC(v_0,\tilde v)$ exiting from $v$ and pointing to $v'$.
 
Since the bound in \S 4.8 is independent on the exact structure
of the contractions, the bound \equ(4.35) can be still obtained,
so that also $\HH\in
C^{(s)}(\TTT^{\ell})$, if $f\in \hat C^{(2+s+{\h\over 2}
\t+\m)}(\TTT^{\ell})$, with $\m>0$.
 
Thus the proof of Theorem 1.4 is complete.

\vskip1.truecm

\centerline{\titolo 5. Comparison with the one-dimensional}
 
\centerline{\titolo Schr\"odinger equation in a quasi periodic potential}
\*\numsec=5\numfor=1
 
\0{\bf 5.1.} From Theorem 1.4,
one could deduce the existence of Bloch waves for the
one-dimensional Schr\"odinger equation with a potential
belonging to a certain class of non analytic quasi periodic functions,
and one could be tempted to compare the result with [Pa],
where the existence of Bloch waves is proven with the
Moser-Nash techniques for quasi periodic potentials
having $p>2(\ell+1)$ continuous derivatives
(if $\ell$ is the dimension of the frequency vector
of the quasi periodic potential and $\t$ is supposed
to be $\t>\ell-1$), with no other restriction
on the potential regularity.
However, in order to perform a meaningful comparison
between the two results, one has to consider carefully
the exact form of the interaction potential.
 
\*
 
\0{\bf 5.2.} The problem studied in [DS,R,P] is
the Schr\"odinger equation
$$ \Big[ -{d^2\over dx^2} + \e V(x) \Big] \psi(x) = E\psi(x) \;
\Eq(5.1) $$
where $V(x)$ is a quasi periodic function of the form
$$ V(x) = \sum_{\nn\in \zzz^{\ell-1} } e^{i\oo\cdot\nn x} \, V_\nn
\; , \Eq(5.2) $$
with $\oo\in\RRR^{\ell-1}$ satisfying a Diophantine condition.
 
The problem to find eigenvalues and eigenfunctions of \equ(5.1)
can be easily seen to be equivalent to the problem to solve
the equations of motion of the classical mechanics system
described by the Hamiltonian
$$ \HHH = {p^2\over2} + \oo\cdot\BB + {q^2\over2}
\Big[ E - \e V(\bb) \Big] \; , \Eq(5.3) $$
with $(p,q)\in\RRR^2$ and $(\BB,\bb)\in \RRR^{\ell-1}\times
\TTT^{\ell-1}$. In fact the evolution equation for
the coordinate $q$ is nothing but the eigenvalue equation \equ(5.1).
 
Then it is possible to introduce a canonical transformation
$\CC\!:(p,q)\to(A_1,\a_1)$, [G2], such that
$$ \HHH = \sqrt{E} A_1 + \oo\cdot\BB +
\e f(\a_1,\bb) \; , \qquad
f(\a_1,\bb) = - {A_1 \over \sqrt{E} } \left( \sin^2\a_1
\right) V(\bb) \; , \Eq(5.4) $$
which can be reduced to the form \equ(1.3), with
$\AA=(A_1,\BB)\in\RRR^{\ell}$, $\aa=(\a_1,\bb)\in \TTT^{\ell}$, and
$\e\ff(\aa)=(f(\aa),0,\ldots,0)$.  For the proof of such an assertion,
we refer to [G2].
 
Then the equations of motions for $\bb$ gives $\bb(t) =\bb_0+\oo t $,
and the derivatives whose number can grow up indefinitely are those
acting on the $\a_1$ variable: {\it however the perturbation is always
analytic in $\a_1$}.
 
Thus the reason why in Theorem 1.4 we needed much stronger assumptions
on the interaction potential, compared with \equ(5.4), is simply that
the one-dimensional Schr\"odinger equation can be reduced to a classical
mechanics problem with Hamiltonian of the form \equ(1.1), but the
interaction term depends analytically on $\a_1$, independently on the
regularity of the quasi periodic potential.
 
In this case the existence of the counterterm can be proven without
exploiting the ultraviolet cancellations, and the infrared cancellations
are sufficient to give the convergence of the perturbative series,
provided the quasi periodic potential is so regular to guarantee the
summability on the Fourier components in the perturbative series:
the analysis in \S 3,4 gives $p>1+3\t$, see Appendix A3 for details.
 
Then, if $\t>\ell-1$, one has $p>3\ell-2$.  With respect to [Pa], the
result is weaker for $\ell\ge5$, but stronger for $\ell\le3$ (and
equivalent for $\ell=4$).  Nevertheless, as the result in [Pa] has been
obtained by using the Moser-Nash techniques for KAM theory, and it is
known that the class of differentiability of the perturbations of
integrable systems can be raised from the Moser result $p>2(\ell+1)$ to
(the perhaps optimal) $p>2\ell$, [P], {\it then one can conjecture that
also for the Schr\"odinger equation the ideas in} [P] {\it would lead to
$p>2\ell$}. Therefore our result $p>3\ell-2$ can be considered, for low
$\ell$, an improvement of [Pa].
 
Furthermore it is important to note also that with the techniques
described in the present paper no other regularity condition is required
for the perturbation other than that of being differentiable enough (\ie
no special class of functions as $\hat C^{(p)}(\TTT^{\ell})$ has to be
invoked), in order to obtain {\it analyticity} in the perturbative parameter
of the eigenvalue $E$ and the eigenfunction $\psi(x)$ in \equ(5.1).
 
\*
 
\0{\bf 5.3.} The situation is essentially identical
if one consider the Schr\"odinger equation
$$ \Big[ -{d^2\over dx^2} + U(x) + \e V(x) \Big] \psi(x) = E\psi(x) \;
\Eq(5.5) $$
where $U(x)$ is a periodic potential with frequency $\o_2$ and
$V(x)$ a quasi periodic function of the form
$$ V(x) = \sum_{\nn\in \zzz^{\ell-2} } e^{i\oo\cdot\nn x} \, V_\nn
\; , \Eq(5.6) $$
with $\oo\in\RRR^{\ell-2}$, such that $\o_2$ and $\oo$
satisfy a Diophantine condition.
 
The Hamiltonian of the corresponding classical mechanics problem is
$$ \HHH = {p^2\over2} + \o_2 B_2 + \oo\cdot\BB + {q^2\over2}
\Big[ E - U(\b_2) - \e V(\bb) \Big] \; , \Eq(5.7) $$
with $(p,q)\in\RRR^2$, $(B_2,\b_2)\in \RRR^{1}\times \TTT^{1}$,
and $(\BB,\bb)\in \RRR^{\ell-1}\times \TTT^{\ell-1}$.
If $\e=0$, the Hamiltonian is integrable,
[G2,C], so that \equ(5.7) becomes
$$ \eqalign{
& \HHH = \o_{01} A_1 + \o_{02} A_2 + \oo\cdot\BB +
\e f(\a_1,\a_2, \bb) \; , \cr
& f(\a_1,\a_2, \bb) = - A_1\,G(\a_0,\a_1)\,V(\bb) \; , \cr} \Eq(5.8) $$
where $G(\a_1,\a_2)$ is a function which depends analytically
on $\a_0$, [C], \S V,VI, independently on the regularity
of $U$ and $V$ in \equ(5.5). The fact that the interaction
is proportional only to $A_1$ (\ie independent on the
other action variables) implies that the equations
of motion for $\a_1$ and $\bb$ can be trivially integrated and give
$\a_2=\a_{20}+\o_{02}t$ and $\b_j=\b_{j0}+\o_{0j}t$, $2\le j\le\ell$.
Then we can reason as in \S 5.2,
and the same conclusions hold.

\vskip1.truecm

\centerline{\titolo Appendix A1. Graphs and graph rules}
\*\numsec=1\numfor=1
 
\0We lay down one after the other, on a plane, $k$ pairwise distinct unit
segments oriented from one extreme to the other: respectively the {\sl
initial point} and the {\sl endpoint} of the oriented segment.
The oriented segment will also be called {\sl arrow}, {\sl branch} or
{\sl line}. The segments are supposed to be numbered from $1$ to $k$.
 
The rule is that after laying down the first segment, the {\sl root
branch}, with the endpoint at the origin and otherwise arbitrarily, the
others are laid down one after the other by attaching an endpoint of a
new branch to an initial point of an old one and by leaving free the
new branch initial point.  The set of initial points of the object thus
constructed will be called the set of the graph {\sl nodes} or {\sl
vertices}.  A graph of {\sl order} $k$ is therefore a partially ordered
set of $k$ nodes with top point the endpoint of the root branch, also
called the {\sl root} (which is not a node); in general
there will be several ``bottom nodes" (at most $k-1$).
 
We denote by $\le$ the ordering relation, and say that two nodes $v$, $w$
are ``comparable'' if $v<w$ or $w<v$.
 
With each graph node $v$ we associate an {\sl external momentum} or
{\sl mode} which is simply an integer component vector $\nn_v\ne{\bf 0}$;
with the root of the graph (which is not regarded as a node) we
associate a label $j=1,\ldots,\ell$.
 
For each node $v$, we denote by $v'$ the node immediately following $v$
and by $\l_v\=v'v$ the branch connecting $v$ to $v'$, ($v$ will be the
initial point and $v'$ the endpoint of $\l_v$). If $v$ is the node
immediately preceding the root $r$ ({\sl highest node}) then we shall
write $v'=r$, for uniformity of notation (recall that $r$ is not a node).
 
We consider ``comparable'' two lines $\l_v$, $\l_w$, if $v$, $w$ are such.
 
If $p_v$ is the number of branches entering the
node $v$, then each of the $p_v$ branches can be thought as the root
branch of a {\sl subgraph} having root at $v$: the subgraph is uniquely
determined by $v$ and one of the $p_v$ nodes $w$ immediately preceding
$v$. Hence if $w'=v$ it can be denoted $\th_{vw}$.
 
We shall call {\sl equivalent} graphs which can be overlapped by\\
(1) changing the angles between branches emerging from the
same node, or\\
(2) permuting the subgraphs entering into a node $v$
in such a way that all the labels match.
 
The number of (non equivalent numbered) graphs with $k$ branches
is bounded by $4^k k!$, [HP].

\vskip1.truecm

\centerline{\titolo Appendix A2. Proof of Lemma 3.4}
\*\numsec=2\numfor=1
 
\0We consider all the graphs we obtain by detaching from each resonance
the subgraph with root $v_V^b$, if $v_V^b$ is the node in which the
resonance line $\l_V$ enters, then reattaching it to all the nodes
$w\in V$.  We call this set of contributions a {\sl resonance family}. 
If one set $\z\=\oo_0\cdot\nn_{\l_V}=0$, no propagator changes inside
the resonance, and the only effect of the above operation is that in
the factor
$$ \Big[ (\nn_{v_V^0}\cdot\ff_{\nn_{v_V^a}})\,
(\nn_{v_V^b}\cdot\ff_{\nn_{v_V^1}})\,
{ \ch(2^{-n_\l}\oo_0\cdot\nn_{\l_V}) \over
(\oo_0\cdot\nn_{\l_V})^2} \Big] \Eqa(A2.1) $$
appearing in \equ(3.3) the external momentum $\nn_{v_V^b}$ assumes
successively the values $\nn_w$, $w\in V$.  In this way, by summing
over all the trees belonging to a given resonance family, we build a
quantity proportional to $\sum_{w\in V} \nn_w=\V0$, by Definition 3.3
of real resonance.

\vskip1.truecm

\centerline{\titolo Appendix A3. Regularity of the potential}
 
\centerline{\titolo for the Schr\"odinger equation in a quasi periodic
potential}
 
\*\numsec=3\numfor=1
 
\0The equations of motion of the Hamiltonian system \equ(5.4), once a
counterterm $A_1N_1(\e)$ has been added, give for the angle variables
$$ \eqalign{
{d\a_1\over dt} & = \sqrt{E} - \e {1\over\sqrt{E}}(\sin^2\a_1)V(\oo_0 t)
+ N_1(\e)\; , \cr
{d\bb\over dt} & = \oo_0 \; , \cr} \Eqa(A3.1) $$
which can be discussed as in \S 2. A formula in terms of graphs
\equ(2.6) can be still obtained, where $\hbox{Val}(\th)$,
defined in \equ(2.4), becomes
$$ \hbox{Val}(\th) = \prod_{v<r}
\Big( - {1\over\sqrt{E}} \Big)
\, { \n_{v'}s_{\n_v}V_{\nn_v}
\over \sqrt{E}\n_{\l_v} + \oo\cdot\nn_{\l_v} } \; , \Eqa(A3.2) $$
where $\nn_v, \nn_{\l_v}\in\ZZZ^{\ell-1}$
and $\n_v\in\{-2,0,2\}$ $\forall v\in\th$, and we decomposed
$$ V(\bb)=\sum_{\nn\in\zzz^{\ell-1}} e^{i\nn\cdot\bb}\,
V_{\nn} \; , \qquad \sin^2\a_1=\sum_{\n=0,\pm2} e^{i\n\a_1}\,s_\n \;
\Eqa(A3.3) $$
Then one sees that no problem arises from the numerators, (as the only
appearing external momenta are of the form $\n_v$, and $\sin^2\a_1$ is
a trigonometric polynomial in $\a_1$), while the small divisors can be
dealt with through Lemma 3.4.  This gives a factor $|\nn_v|^{3\t}$ per
node, so that summability on the Fourier labels requires at least
$V\in C^{(p)}(\TTT^{\ell-1})$, $p>3\t$.
 
\*
 
The equations of motion for the action variables give
$$ \eqalign{
{d A_1\over dt} & = \e A_1 {\dpr\sin^2\a_1\over\dpr\a_1}\,
V(\oo_0 t) \; , \cr
{d \BB\over dt} & = \e A_1 \sin^2\a_1\,
{\dpr V(\bb) \over\dpr\bb}\Big|_{\bb=\oo_0 t} \; , \cr} \Eqa(A3.4) $$
so that we can reason as above, with the only
difference that the highest node of the graph $v_0$ has a
factor $\nn_{v_0}$ which requires, to guarantee
the summability on $\nn_{v_0}$,
$V\in C^{(p)}(\TTT^{\ell-1})$, with $p>3\t+1$.
 
\*
 
Then one has to require at least $V\in C^{(p)}(\TTT^{\ell-1})$,
$p>3\t+1$, in order to have $\a_1\in C^{(1)}(\TTT^{1})$,
as it has to be for the Schr\"rodinger equation
\equ(5.1) to be meaningful, if one recalls that
(1) the wave function $\psi(x)$ solving \equ(5.1)
has to be of class $C^{(1)}$ for $V\in C^{(0)}$,
and (2) $\psi(x)=q(x)$, where $q$ is the variable
related with $\a_1$ by the canonical transformation $\CC$
defined before \equ(5.4).

\vskip1.truecm

\centerline{\titolo Appendix A4. Comparison
between Moser's counterterms theorem}
 
\centerline{\titolo and the counterterms conjecture in [G1].}
\*\numsec=4\numfor=1
 
\0{\bf A4.1.} In [M1] a perturbation theory for quasi-periodic
solutions of a nonlinear system of ordinary differential equations
is developed. Up to a (trivial) coordinate transformation,
the system can be written in the form
$$ \eqalign{
{d\xx\over dt} & = \oo + \e \ff(\xx,\yy;\e) \; , \cr
{d\yy\over dt} & = \Omega \yy + \e \gg(\xx,\yy;\e) \; , \cr} \Eqa(A4.1) $$
where $\xx\=(x_1,\ldots,x_n)\in\RRR^n$,
$\yy\=(y_1,\ldots,y_m)\in\RRR^m$, $\oo\in\RRR^n$,
$\Omega$ is a constant $m\times m$ matrix with eigenvalues
$\Omega_1,\ldots, \Omega_m$,
and $\ff$ and $\gg$ are functions
with period $2\p$ in $x_1,\ldots,x_n$ and analytic
in $\xx,\yy$ and $\e$ (in suitable domains).
 
If the characteristic numbers $\o_1,\ldots,\o_n,\O_1,\ldots,\O_n$
verify the generalized Diophantine condition
$$ C_0 \Big| i \sum_{j=1}^n \n_j\o_j + \sum_{i=1}^{m}
\m_i\O_i \Big| \ge \big( |\nn|^{\t}+1 \big)^{-1} \; ,
\Eqa(A4.2) $$
with $\nn\=(\n_1,\ldots,\n_m)\in\ZZZ^n$,
$|\nn|=\sum_{j=1}^n\n_j$, and $(\m_1,\ldots,\m_m)\in\ZZZ^m$
then there exists a unique analytic vector valued functions
$\ll(\e)$ and $\mm(\e)$ and a unique analytic matrix
valued function $M(\e)$ such that the modified system
$$ \eqalign{
{d\xx\over dt} & = \oo + \e \ff(\xx,\yy;\e) + \ll(\e) \; , \cr
{d\yy\over dt} & = \Omega \yy + \e \gg(\xx,\yy;\e) +
\mm(\e) + M(\e)\yy \; , \cr} \Eqa(A4.3) $$
admits a quasi periodic solution with the same
characteristic number as the unperturbed one, [M1], Theorem 1.
 
\*
 
\0{\bf A4.2.} Let us consider the case in which
$m=n=\ell$, $\O=0$, and there exists a function
$\HHH_0=\oo\cdot\yy+\e f(\xx,\yy;\e)$
such that $\ff=\dpr_{\yy}f$ and
$\gg=-\dpr_{\xx}f$. Then the system \equ(A4.1) becomes
the system studied in [GM2], \S 8.
 
Under the same hypotheses, if moreover
$f(\xx,\yy;\e)\=\e\yy\cdot\ff(\xx)$ for some function $\ff$,
\equ(A4.1) and \equ(A4.3) become the equations
of motion of systems described by the
Hamiltonians, respectively, \equ(1.1) and \equ(1.3).
In fact the linearity in the action variables of the term added to the
Hamiltonian $\HHH_0$ in \equ(1.3) leads to a term independent of the
action variables in the equations of motion, \ie $\NN(\e)\=\ll(\e)$,
while the counterterms $\mm(\e)$ and $M(\e)$ are identically vanishing
as a consequence of the symplectic structure of the equations of motion
(as one can argue {\it a posteriori} from Theorem 1.4 in \S 1).
 
In the general case in which the function $f(\xx,\yy;\e)$ appearing in
the Hamiltonian $\HHH_0$ depends arbitrarily (but always analytically)
on $\yy$, the systems studied in [M1] (under the same hypotheses as
above) and in [GM2] are no longer equal to each other, \ie the modified
system \equ(A4.3) is {\it not} the system with Hamiltonian considered
Eq. (1.10) of [GM2], so that Theorem 1.4 in [GM2] can not be reduced to
the results of [M1]: in fact not only there will be no more a trivial
relation between the counterterms $\NN(\e)$ and $\ll(\e)$, but also the
solutions of the equations of motion will be different from each other.
 
Note however that the result following from Moser's theorem
applied to such a system (\ie a Hamiltonian system
with $\O=0$) can be (trivially) reproduced with
our techniques. Also an extension of our techniques
to Hamiltonian systems
(verifying the anisochrony condition) such that $\O\neq 0$\annota{3}{
The anisochronus case with $\O=0$ is simply the KAM theorem.}
could been envisaged: an example in this direction is in
[Ge], where $\0$ has eigenvalues $\O_1=\ldots=\O_{\ell-1}=0$,
$\O_{\ell}=g^2$, and the existence of a counterterm
$M(\e)$ analytic in $\e$ is proven (while $\ll(\e)\=\mm(\e)\=\V0$
again for the symplectic structure of the equations of motion).
\*
\0{\bf Acknowledgements.} This work is part of the research program of
the European Network on: ``Stability and Universality in Classical
Mechanics", \# ERBCHRXCT940460. One of us (G.Ge) acknowledges
finantial support from EC program TMR and is indebted
to IHES for hospitality.

\*
\centerline{\titolo References}
\*
 
\halign{\hbox to 1.2truecm {[\ottorm#]\hss} &
        \vtop{\advance\hsize by -1.25 truecm \noindent#}\cr
 
BGGM& {F. Bonetto, G. Gallavotti, G. Gentile, V. Mastropietro:
Lindstedt series, ultraviolet divergences and  Moser's theorem,
Preprint IHES/P/95/102 (1995). }\cr
C& {L.Chierchia:
Absolutely continuous spectra of quasi periodic Schr\"odinger operators,
{\it J. Math. Phys.} {\bf 28}, 2891--2898 (1987). }\cr
DS& {E.I. Dinaburg, Ya.G. Sinai:
The one dimensional Schr\"odinger equation
with a quasi periodic potential,
{\it Funct. Anal. and its Appl.} {\bf 9}, 279--289 (1975). }\cr
E1& {L.H. Eliasson:
Absolutely convergent series expansions for quasi-periodic motions,
Report 2--88, Department of Mathematics,
University of Stockholm (1988). }\cr
E2& {L.H. Eliasson:
Hamiltonian systems with linear normal
form near an invariant torus, in {\sl Nonlinear Dynamics},
Ed. G. Turchetti, Bologna Conference, 30/5 to 3/6 1988,
World Scientific, Singapore (1989). }\cr
EV& {Ecalle, J., Vallet, B.: Prenormalization, correction,
and linearization of resonant vector fields or diffeomorphisms,
Preprint 95-32, Universit\'e de Paris Sud - Mathematiques, Paris (1995).}\cr
G1& {G. Gallavotti:
A criterion of integrability for perturbed harmonic
oscillators. ``Wick Or\-der\-ing'' of the
perturbations in classical mechanics and
invariance of the frequency spectrum,
{\it Comm. Math. Phys.} {\bf 87}, 365--383 (1982). }\cr
G2& {G. Gallavotti:
Classical Mechanics and Renormalization Group,
Lectures notes at the 1983 Erice summer school,
Ed. A. Wightman \& G. Velo, Plenum, New York (1985). }\cr
G3& {G. Gallavotti:
Quasi-integrable mechanical systems, in {\sl
Critical Phenomena, Random Systems, Gauge Theories}, Les Houches,
Session XLIII (1984), Vol. II, 539--624, Ed. K. Osterwalder \& R. Stora,
North Holland, Amsterdam (1986). }\cr
G4& {G. Gallavotti:
Twistless KAM tori, {\it Comm. Math. Phys.} {\bf 164}, 145--156 (1994). }\cr
Ge& {G. Gentile:
Whiskered tori with prefixed frequencies and Lyapunov spectrum,
{\it Dynam. Stability of Systems} {\bf 10}, 269--308, (1995).}\cr
GM1& {G. Gentile, V. Mastropietro:
KAM theorem revisited,
{\it Physica D} {\bf 90}, 225--234 (1996). }\cr
GM2& {G. Gentile, V. Mastropietro:
Methods for the analysis of the Lindstedt series for KAM tori
and renormalizability in classical mechanics. A review with
some applications, to appear in {\it Rev. Math. Phys.} }\cr
GGM& {G. Gallavotti, G. Gentile, V. Mastropietro: Field theory and KAM
tori, {\it Mathematical Physics Electronic Journal} {\bf 1},
(5): 1--13 (1995). http://mpej@ math. utexas. edu}\cr
 
HP& {F. Harary, E. Palmer: {\sl Graphical enumeration},
Academic Press, New York, 1973. }\cr
M1& {J. Moser:
Convergent series expansions for quasi periodic motions,
{\it Math. Ann.} {\bf 169}, 136-176 (1967). }\cr
M2& {J. Moser:
On the construction of almost periodic solutions for ordinary
differential equations, {\it Proceedings of the
International Conference of functional
analysis and related topics}, Tokyo, 60--67, (1969). }\cr
Pa& {I.S. Parasyuk:
On instability zones of the Schr\"odinger equation with a quasi periodic
potential, Ukr. Mat. Zh. {\bf 30}, 70-78 (1985). }\cr
PF&{L.A. Pastur, A. Figotin:
{\sl Spectra of random and almost-periodic operators},
Grund\-lehren der Mathe\-mati\-schen Wissen\-schaften {\bf 297},
Sprin\-ger, Berlin (1992).}\cr
P& {J. P\"oschel:
\"Uber invarianten Tori on differenzierbaren Hamiltonschen Systemen,
{\it Bonn. Math. Schr.} {\bf 120}, 1-120 (1980). }\cr
R& {H. R\"ussmann:
One dimensional Schr\"odinger equation,
{\it Ann. New York Acad. Sci.} {\bf 357}, 90--107 (1980). }\cr
S& {C.L. Siegel:
Iterations of analytic functions,
{\it Ann. of Math.} {\bf 43}, No.4, 607--612 (1943). }\cr
}
\ciao